\documentclass[aps,prx,longbibliography,twocolumn,showpacs,preprintnumbers,amsmath,amssymb,superscriptaddress]{revtex4-1} 
\usepackage{graphicx}
\usepackage{bm}
\usepackage{psfrag}
\usepackage{subfigure}
\usepackage{color}
\usepackage[normalem]{ulem}
\usepackage[utf8]{inputenc}
\DeclareUnicodeCharacter{2212}{$-$}
\usepackage{tikz}
\usetikzlibrary{decorations.pathreplacing}
\usepackage{pgf}
\usepackage{import}
\usepackage{pgfplots}
\usepackage{hyperref}
\usepackage{import}

\newcommand{\be}{\begin{equation}}
\newcommand{\ee}{\end{equation}}
\newcommand{\ba}{\begin{eqnarray}}
\newcommand{\ea}{\end{eqnarray}}

\newcommand{\ma}[1] {\textcolor{black}{#1}}

\newcommand{\mm}[1]{#1}

\begin{document}

\author{Mario Geiger}
\thanks{These two authors contributed equally.}
\affiliation{Institute of Physics, EPFL, CH-1015 Lausanne, Switzerland}

\author{Stefano Spigler}
\thanks{These two authors contributed equally.}
\affiliation{Institute of Physics, EPFL, CH-1015 Lausanne, Switzerland}

\author{St\'ephane d'Ascoli}
\affiliation{Institut de Physique Th\'eorique, Universit\'e Paris-Saclay, CEA, CNRS, F-91191  Gif-sur-Yvette, France}
\affiliation{Laboratoire de Physique Statistique, \'Ecole Normale Sup\'erieure,
PSL Research University, F-75005 Paris, France}

\author{Levent Sagun}
\affiliation{Institut de Physique Th\'eorique, Universit\'e Paris-Saclay, CEA, CNRS, F-91191  Gif-sur-Yvette, France}
\affiliation{Institute of Physics, EPFL, CH-1015 Lausanne, Switzerland}

\author{Marco Baity-Jesi}
\affiliation{Department of Chemistry, Columbia University, 10027 New York, USA}

\author{Giulio Biroli}
\affiliation{Institut de Physique Th\'eorique, Universit\'e Paris-Saclay, CEA, CNRS, F-91191  Gif-sur-Yvette, France}
\affiliation{Laboratoire de Physique Statistique, \'Ecole Normale Sup\'erieure,
PSL Research University, F-75005 Paris, France}

\author{Matthieu Wyart}
\affiliation{Institute of Physics, EPFL, CH-1015 Lausanne, Switzerland}

\date{\today}



\title{The jamming transition as a paradigm to understand the loss landscape of deep neural networks}

\begin{abstract}
Deep learning has been immensely successful at a variety of tasks, ranging from  classification to artificial intelligence.  Learning corresponds to fitting training data, which is implemented by descending a very high-dimensional loss function. Understanding under which conditions neural networks do not get stuck in poor minima of the loss, and how  the landscape of that loss evolves as depth is increased remains a challenge. Here we predict, and test empirically, an analogy between this landscape and the energy landscape of repulsive  ellipses. We argue that in fully-connected deep networks a phase transition delimits the over- and under-parametrized regimes where fitting can or cannot be achieved. In the vicinity of this transition, properties of the curvature of the minima of the loss (the spectrum of the hessian) are critical. This transition shares direct similarities with the jamming transition by which  particles form a disordered solid as the density is increased\mm{, which also occurs in certain classes of computational optimization and learning problems such as the perceptron}. Our analysis gives a simple explanation as to why poor minima of the loss cannot be encountered in the overparametrized regime. Interestingly,  we observe  that the ability of fully-connected networks to fit random  data is independent of their depth,  an independence  that appears to also hold for real data.
We also study a quantity $\Delta$ which characterizes how well ($\Delta<0$) or badly ($\Delta>0$) a datum is learned. At the critical point it is power-law distributed on several decades, $P_+(\Delta)\sim\Delta^\theta$ for $\Delta>0$ and $P_-(\Delta)\sim(-\Delta)^{-\gamma}$ for $\Delta<0$, with exponents that depend on the choice of activation function. This observation suggests that near the transition the loss landscape has a hierarchical structure and that the learning dynamics is prone to avalanche-like dynamics, with abrupt changes in the set of patterns that are learned.

\end{abstract}


\maketitle

\section{Introduction}
Deep neural networks are now central tools for a variety of tasks including image classification~\cite{Krizhevsky12,Lecun15}, speech recognition~\cite{Hinton12} and the development of artificial intelligence
that can for example master the game of Go beyond human level~\cite{Silver16,Silver17}. 
A neural network represents a (very high-dimensional) function $f$ that depends on a large number of parameters $N$~\cite{Lecun15}.  These parameters are learned so as to correctly classify $P$ training data by minimizing some loss function ${\cal L}$, generally via  stochastic gradient descent (a kind of noisy version of gradient descent). There is great flexibility in the network architecture, loss function and minimization protocol one can use. These features are ultimately selected to optimize the classification of previously unseen data, or  {\it generalization}. Although the current progress in designing~\cite{Lecun95,He16}
and training~\cite{Ioffe15} networks that generalize well is undeniable, it remains mostly empirical. A general theory explaining and fostering this success is lacking, and central questions remain to be clarified. First, since the loss function is generally not convex, why doesn't  the learning dynamics get stuck in poorly performing minima with high loss? In other words, under which conditions can one guarantee that training data are well fitted? Second, what are the benefits of deeper networks? \mm{On the one hand } it is often argued, and proved in some cases, that the advantage of deep networks  stems from their enhanced expressive power, i.e.\ their ability to build complex functions with a much smaller number of parameters than needed for shallow networks~\cite{Montufar14,Bianchini14,Raghu16,eldan2016power,lee2017ability}. Indeed if deep networks are able to fit data with less parameters, then they are likely to generalize better.  \mm{On the other hand,} one can handcraft neural networks that fit even structure-less, random data with a rather small number of parameters $N\sim P$~\cite{Gardner88,Monasson95,Zhang16,Baum88}. \mm{These results for the static capacity of networks appear to be independent of depth \cite{Zhang16,Baum88}}. Yet, it is unclear whether such parsimonious solutions can be found \mm{dynamically} in practice simply by descending the loss function, and whether depth can help finding them. More generally, how is the loss landscape affected by depth?

Complex physical systems with non-convex energy landscapes featuring an exponentially large number of local minima are called glassy \cite{reviewBB}. Does the landscape of deep learning fall into a known class of glassy systems? Along this line, an analogy between deep networks and mean-field glasses ($p$-spins)  has been proposed~\cite{Choromanska15}, in which the learning dynamics is expected to get stuck in the highest minima of the loss, which are the most abundant. Yet, several numerical  and rigorous works~\cite{Freeman16,Hoffer17,Soudry2016,Cooper18} (the latter focusing on shallow and very overparametrized networks) suggest a different landscape geometry where the loss function is characterized by a connected level set. Furthermore,  studies of the Hessian of the loss  function~\cite{Sagun16,sagun2017empirical,Ballard17} and of the learning dynamics~\cite{Lipton16,Baity18} support that the landscape is characterized by an abundance of flat directions, even near its bottom, at odds with traditional glassy systems. 

In the last decade several works have unveiled an analogy between the physical phenomenon of jamming~\cite{Wyart05b,Liu10} and phase transitions taking place in certain classes of computational optimization and learning problems~\cite{Krzakala07,Zdeborova07,Franz17}, in particular the perceptron ~\cite{Franz16,Franz17} --- the simplest neural network performing linear classification. 
In this work we push this analogy  further and show that the geometry of the training loss landscape and the training dynamics of fully connected deep neural networks
\mm{is affected by a}  jamming transition \mm{similar to that} of repulsive ellipses~\cite{Liu10}. As illustrated in Fig.~\ref{fig1}, jamming occurs in packings of particles interacting through a finite-range potential ${\cal U}$, when the particle density $\phi$ reaches some critical value $\phi_c$. At that point, particles can no longer be accommodated without touching each other and the system becomes a solid with singular landscape properties, embodied for example in the spectrum of the Hessian of ${\cal U}$~\cite{Wyart05a,Silbert05}, that at the transition displays many (almost) flat directions. Particles of different shapes, such as spheres and ellipses, can lead to different jamming scenarios~\cite{Donev04a,Mailman09,Zeravcic09,Brito18a}.
\begin{figure}[bht]
    \centering
    \def\svgwidth{0.6\columnwidth}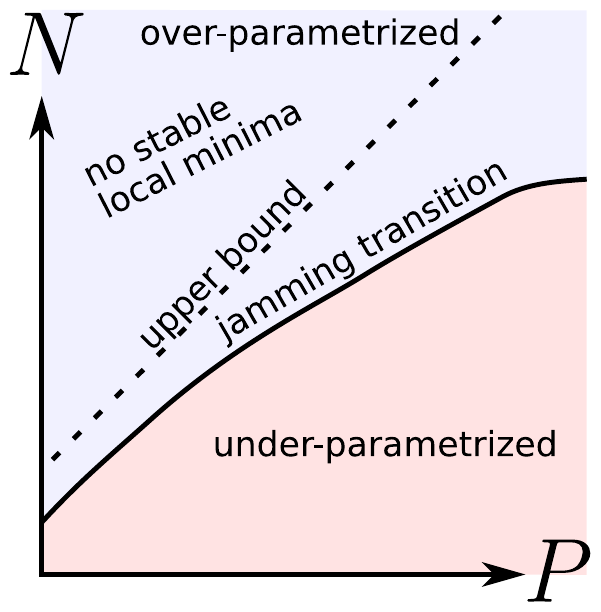
    \caption{\small $N$: degrees of freedom, $P$: training examples.}
\end{figure}

Here we show that for two commonly used loss functions (cross-entropy and quadratic hinge), fully-connected deep networks undergo a jamming transition too, below which all data are correctly fitted and above which they are not, both for real data (images) and random data~\footnote{This transition influences the generalization properties of deep networks, too. This has been observed, for instance, in~\cite{advani2017high,spigler2018jamming}, and studied by the authors in~\cite{geiger2019scaling} (preprint).}. In both cases the transition appears to be solely controlled by the number of parameters of the network $N$, independently of depth. For random data, the transition takes place as the quantity $P/N$ increases toward some critical value $P/N^*$. For the hinge loss, using results from the jamming literature we argue that $P/N^*\geq C_0$ where $C_0$ is a constant that we can measure \emph{a posteriori} once learning took place. To hold, this result requires the network output to remain sensitive to all its weights during training, as we observe empirically in the examples we study. This view supports that the dynamics cannot get stuck in poor minima in the over-parametrized regime where networks tend to operate, because there are not enough constraints to form minima in that regime. We also find that the jamming transition is sharp and the landscape appears to fall in the same universality class independently of depth (as long as at least one hidden layer is present). 
Differently from the (non-convex) perceptron, that was  proven to lie in the same universality class as spherical particles~\cite{Franz16,Franz17}, we show that deep networks instead jam in a manner similar to ellipses. From our analysis we deduce the singular properties of the spectrum of the Hessian of the loss, which indeed must display many flat directions. We find empirically that other key quantities (the fraction of data which are almost correctly or almost incorrectly classified) display power-law behaviours on several decades, with new exponents. In glassy systems, such power-laws reveal properties that cannot be reached by studying the Hessian, in particular the fact that the dynamics occurs via broadly distributed avalanches~\cite{Wyart12,Muller14,Franz17b}, indicative of a hierarchical organization of the landscape~\cite{Charbonneau14}. This observation thus suggests that these properties also characterize deep networks near the transition. \mm{Note that} in this work we focus on training and the ability of deep neural networks to fit a dataset. The implications and the relations with generalization \ma{between jamming and generalization} are investigated in~\cite{spigler2018jamming,geiger2019scaling}. 


%
%
%
%
%
%

\begin{figure*}
    \centering
    \setlength{\unitlength}{0.1\textwidth}
    \begin{picture}(10,4)
    \put(0,2){\def\svgwidth{0.24\textwidth}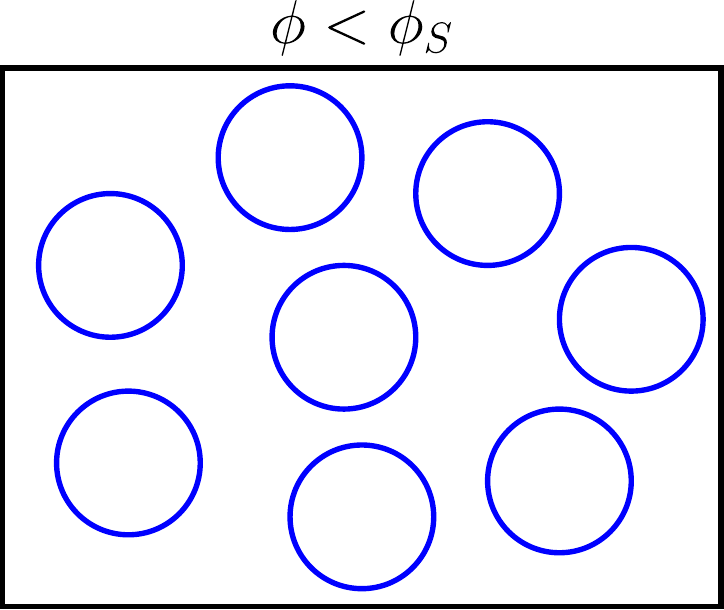}\put(2,3.9){(a)}
    \put(2.5,2){\def\svgwidth{0.24\textwidth}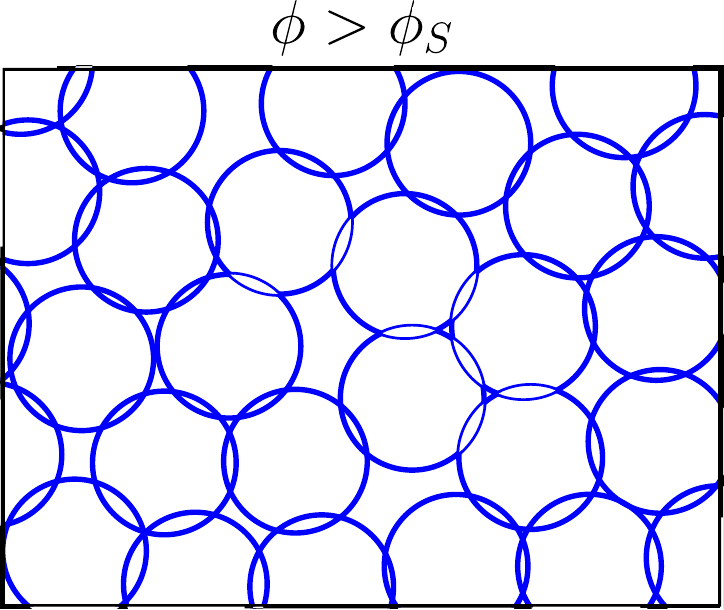}\put(4.5,3.9){(b)}
    \put(5,2){\def\svgwidth{0.24\textwidth}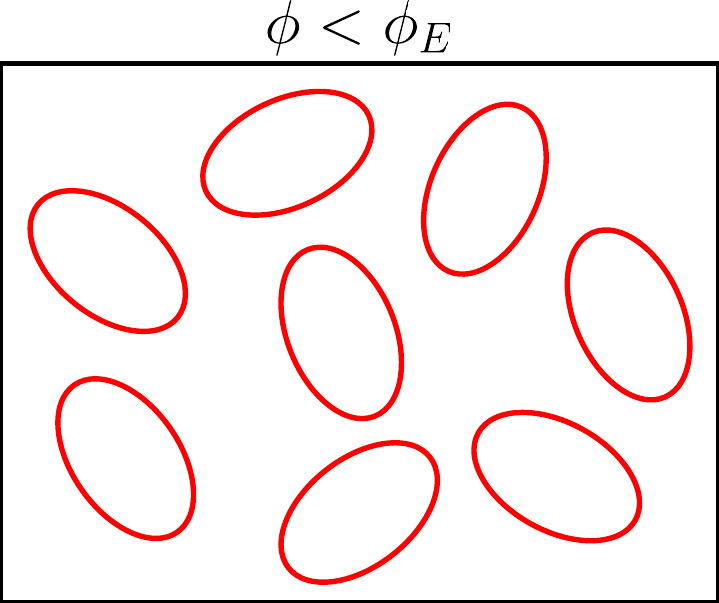}\put(7,3.9){(c)}
    \put(7.5,2){\def\svgwidth{0.24\textwidth}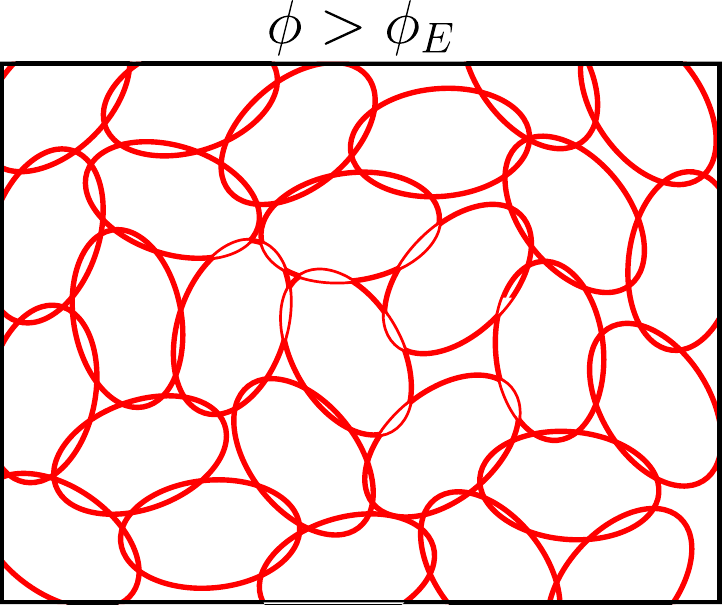}\put(9.5,3.9){(d)}
    \put(0,-0.05){\def\svgwidth{0.24\textwidth}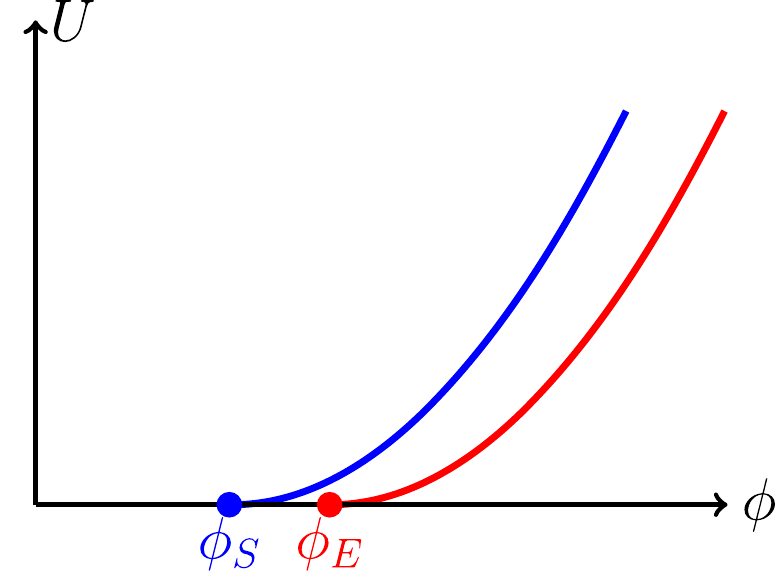}\put(2,1.6){(e)}
    \put(2.5,0.1){\def\svgwidth{0.24\textwidth}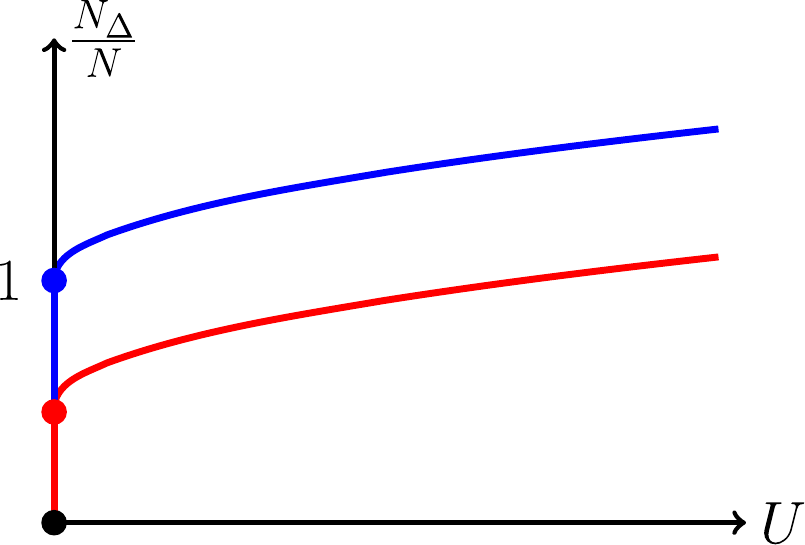}\put(4.5,1.6){(f)}
    \put(5,0){\def\svgwidth{0.24\textwidth}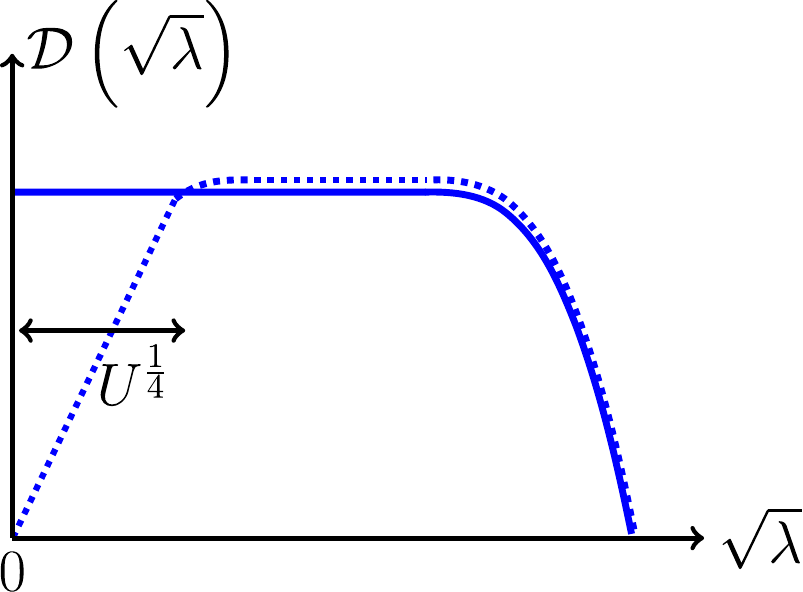}\put(7,1.6){(g)}
    \put(7.5,0){\def\svgwidth{0.24\textwidth}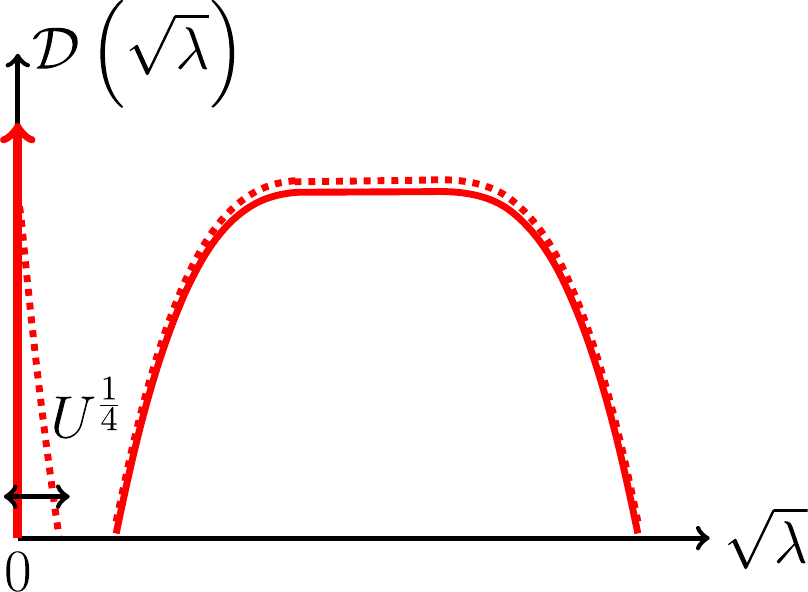}\put(9.5,1.6){(h)}
    \end{picture}
    \caption{Sketch of the jamming transition for repulsive spheres and ellipses. (a,b,c,d) Both systems transition from a fluid to a solid as the density passes some threshold, noted $\phi_S$ for spheres and $\phi_E$ for ellipses. (e) For denser packings, the potential energy ${\cal U}$ becomes finite. (f) The ratio $N_\Delta/N$ between the number of  particles in contact $N_\Delta$ (corresponding to unsatisfied constraints) and the number of degrees of freedom $N$ jumps discontinuously to a finite value, which is unity for spheres but smaller for ellipses. (g,h) This difference has dramatic consequence on the energy landscape, in particular on the spectrum of the Hessian. In both cases, the spectrum becomes non-zero at jamming, but it displays a delta function with finite weight for ellipses (indicating strictly flat directions), followed by a gap with no eigenvalues, followed by a continuous spectrum (h, full line). For spheres, there is no delta function nor gap (g, full line). As one enters the jammed phase, in both cases a characteristic scale $\lambda\sim \sqrt{{\cal U}}$ appears in the spectrum (g and h, dotted lines).\label{fig1}}
\end{figure*}
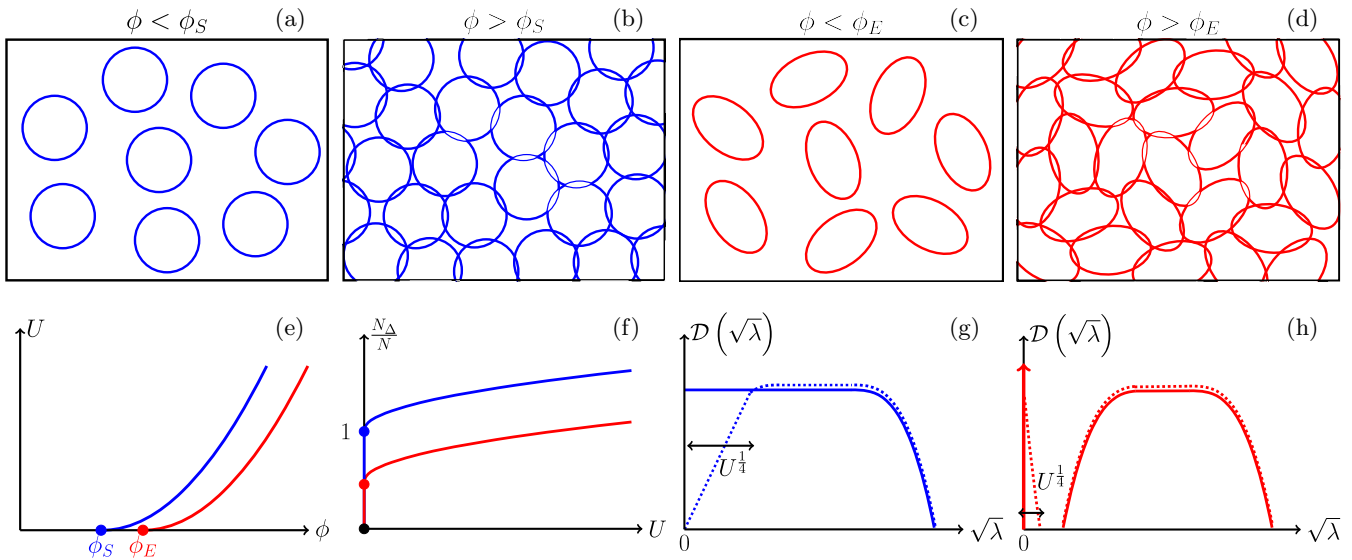

\section{Analogy between jamming and deep learning}
\label{analogy}

\subsection{Jamming}
Understanding the energy landscape --- in particular the properties of the Hessian, referred to as vibrational properties in this context --- in disordered systems of interacting particles is a long-standing and practically important problem~\cite{Phillips81}. It was realized that for purely repulsive, finite-range  particles, such properties are singular near the jamming transition where the system becomes a solid~\cite{Silbert05,Wyart05a}, allowing one to develop and test theories for the vibrations of glasses, that turn out to apply in a broader class of systems where the interactions do not necessarily have finite range~\cite{Wyart05b}. Here we shall follow the same strategy for deep networks, where the role of the ``interaction potential'' is played by the choice of loss function. Finite-range interactions are mimicked by the \emph{hinge loss}, for which we predict a sharp transition when going from an overparametrized to an underparametrized regime. At the transition, the Hessian is singular and displays an abundance of low-energy modes. For other types of losses --- such as for the commonly used cross-entropy loss defined below --- the transition exists but its effects on the spectrum are expected to be less sharp (see discussion below).

We start by recalling some results on the jamming transition. 
We will first discuss the case of spherical particles, since it has been studied thoroughly and is easier to formalize. The behavior of elliptical particles will be discussed later on. Consider spheres of radius $R$ at positions $\{ {\bf r}_i \}$, corresponding to a total number of degrees of freedom $\tilde N$. We denote by $r_{ij}=|\!|{\bf r}_i - {\bf r}_j|\!|$ the distance between particles $i$ and $j$, and define their overlap $\Delta_{ij}=2R-r_{ij}$.  Two particles are said to be in contact if $\Delta_{ij}>0$, and $N_\Delta$ denotes the number of such contacts. We label by $\mu$ all the possible pairs of particles $(ij)$ and by $m$ the sets of contacts. We consider the following potential energy:
\be
\label{1}
{\cal U}=\sum_{\mu \in m} \frac{1}{2}  \Delta_\mu^2.
\ee
We denote by $N$ the effective number of degrees of freedom which affect the variables $\Delta_\mu$. It is in general smaller than $\tilde N$ because of (i)  global translations or rotations of the system and (ii) ``rattlers'', i.e.\ particles which make no contact with the others, whose degrees of freedom are irrelevant as far as the solid phase is concerned.

As the jamming transition is approached from above (large density $\phi$), ${\cal U}\rightarrow 0$ as sketched in Fig.~\ref{fig1},  implying that $\Delta_\mu  \rightarrow 0$ $ \forall \mu \in m$. As argued in
\cite{Tkachenko99}, for each $\mu \in m$ the constraint $\Delta_\mu  = 0$ defines a manifold of dimension $N-1$. Satisfying $N_\Delta$ such equations thus generically leads to a manifold of solutions of dimension $N-N_\Delta$. Imposing that  solutions exist thus implies that, at jamming, one has
\be
\label{2}
N_\Delta \leq N\,.
\ee
Note that this argument implicitly assumes that  the $N_\Delta$ constraints are independent. In disordered systems this assumption is generally correct in practice, but it may break down if  symmetries are present, which is the case e.g.\ in  crystals where Eq.~(\ref{2}) can be violated.

An opposite bound can  be obtained for spheres by considerations of stability, by imposing that in a stable minimum the Hessian must be positive definite~\cite{Wyart05a}.
The Hessian is an $N\times N$ matrix which can be written as~\footnote{A similar decomposition has been used in \cite{pennington2017geometry, martens2014new,sagun2017empirical}.}
\be
\label{3}
{\cal H}_U=\sum_{\mu \in m} \nabla\Delta_\mu \otimes \nabla\Delta_\mu + \sum_{\mu \in m} \Delta_\mu \nabla\otimes\nabla \Delta_\mu \equiv {\cal H}_0 + {\cal H}_p\,,
\ee
where ${\cal H}_0$ and ${\cal H}_p$ correspond to the first and second sum, respectively. ${\cal H}_0$ is positive semi-definite, since it is the sum of $N_\Delta$ matrices of rank unity; thus ${\rm rk}({\cal H}_0) \leq N_\Delta$, implying that the kernel of ${\cal H}_0$ is at least of dimension $N-N_\Delta$. On the other hand for \emph{spheres} --- but not for ellipses, and this will have major consequences --- ${\cal H}_p$ is negative definite, which simply stems from the fact that the second-order contribution  of the displacement to the distance between two points is always positive - a straightforward application of the Pythagoras theorem. It is easy to show~\cite{Wyart05a} that any non-zero vector $|u\rangle$ belonging to the kernel of ${\cal H}_0$  must satisfy $\langle u|{\cal H}_U|u\rangle=\langle u|{\cal H}_p|u\rangle<0$ \footnote{Once again, this statement is true except for global translation or rotation of the systems, whose number however is fixed in the large $N$ limit and disappears when the ratio $N_\Delta/N$ is considered.}. Thus stability requires that ${\rm rk}({\cal H}_0)=N$, implying that $N_\Delta \geq N$. Together with Eq.~(\ref{2}) that leads to  $N_\Delta = N$: as spheres jam  the number of degrees of freedom and the number of constraints (stemming from contacts) are equal, as empirically observed~\cite{Ohern03}.  This property is often called {\it isostaticity}: when it holds, mean-field arguments~\cite{DeGiuli14,Yan16,Franz15} predict that the density of vibrational modes $D(\sqrt{\lambda})$ displays a plateau up to vanishingly small $\lambda$, as observed numerically~\cite{Silbert05,Wyart05a} and sketched in Fig.~\ref{fig1}G.

\begin{figure*}[ht]
    \centering
    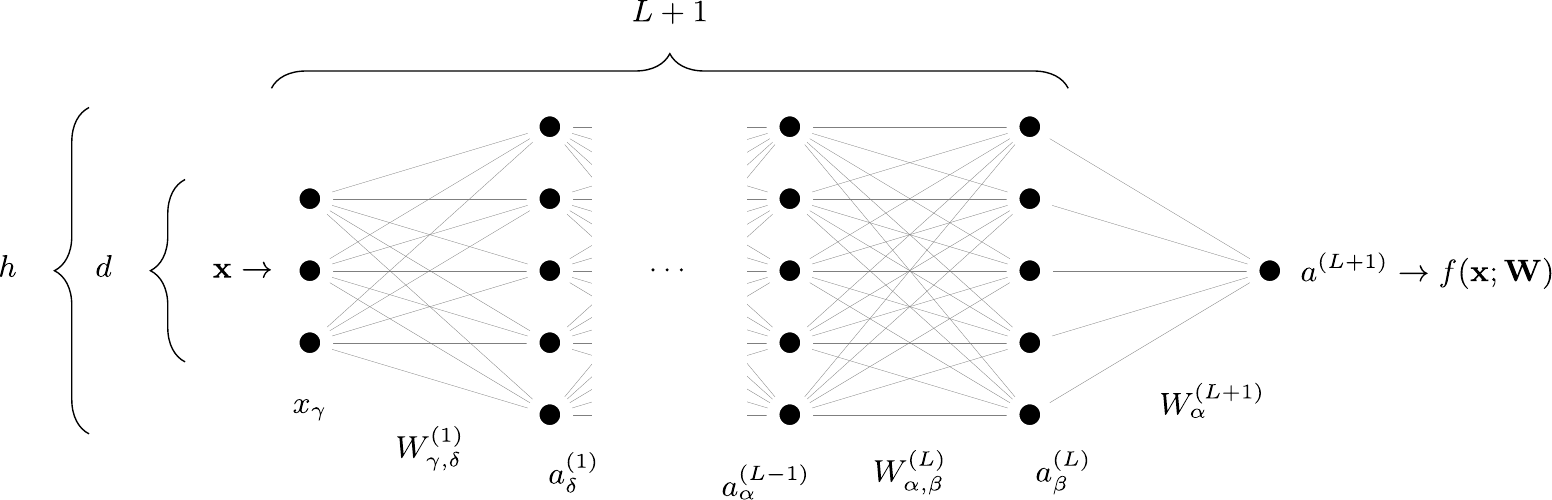
    \caption{Architecture of a fully-connected network with $L$ hidden layers of constant size $h$. Points indicate neurons, connections between them are characterized by a weight. Biases are not represented here. \label{fig:network_architecture}}
\end{figure*}

However, for \emph{ellipses}~\cite{Donev04a} (and as we shall see, for deep networks), this argument breaks down because ${\cal H}_p$ is not negative definite. Whether such a matrix has positive eigenvalues or not plays a role of utmost importance in the selection of the universality class of the jamming transition, and it has major consequences on the singularity of the landscape, as it can be evinced from the spectrum of the Hessian matrix. Indeed,  for ellipses stability and jamming can, and generically do, occur at:
\be
N_\Delta/N<1,
\ee
a situation that is referred to as {\it hypostatic}. The density of vibrational modes at jamming must then display a delta function in zero of magnitude $1- N_\Delta/N$, corresponding to the kernel of ${\cal H}_0$  (${\cal H}_p$ vanishes at jamming since $\Delta_\mu\rightarrow 0$  $ \forall \mu \in m$). Mean-field arguments applied to  hypostatic materials~\cite{During13,Brito18a} predict that at larger $\lambda$, the spectrum presents a gap before becoming continuous again, as sketched in Fig.~\ref{fig1}H. Away from jamming the effects of ${\cal H}_p$ kick in and broaden the delta function by an amount proportional to  the typical value of the overlap $\Delta \sim \sqrt{\cal U}$, as sketched in Fig.~\ref{fig1}H.

We now show that even in the hypostatic case, stability can be constraining. Let us denote by $E_-$ the vector space spanned by the negative eigenvalues of ${\cal H}_p$, whose dimension   very close to jamming  is denoted $N_-$. Stability then imposes that the intersection of  the kernel of ${\cal H}_0$ and $E_-$ is zero, which is possible only if
\be
\label{4}
N_\Delta \geq N_-.
\ee

Finally, another key structural property of the jamming transition is contained in the distribution $P_+(\Delta)$ of \emph{positive overlaps}, sometimes referred to as \emph{forces} (the force between two particles is $\Delta$ when $\Delta>0$), and the distribution $P_-(\Delta)$ of \emph{gaps} ($\Delta<0$) between particles. It was shown that even if  a packing of spherical particles is linearly stable, paths in the phase space that lower the energy are easily found unless both distributions are critical, with $P_+(\Delta)\sim\Delta^\theta$ and $P_-(\Delta)\sim(-\Delta)^{-\gamma}$, with $\gamma\geq (1-\theta)/2$~\cite{Wyart12,Lerner13a}, as numerically confirmed in~\cite{Lerner12,Charbonneau12}.
For a broad class of dynamics, this bound must be saturated~\cite{Muller14}, a scenario referred to as {\it marginal stability} which implies that the dynamics proceeds via power-law distributed events (called avalanches) in which the set of constraints change. Calculations in infinite dimensions~\cite{Charbonneau14,Charbonneau14a} showed that marginal stability is associated with a hierarchical organization of minima of the energy (a phenomenon referred to as \emph{replica symmetry breaking}~\cite{Mezard87}), and exponents were found to follow $\gamma=0.41269\ldots$ and $\theta=0.42311\ldots$ which appear accurate even in finite dimensions~\cite{Lerner13a,Charbonneau15}.\vspace{1em}

\begin{table*}[]
    \centering
    \begin{tabular}{|ccc|}
         \hline
         Particles & vs & Neural networks \\ \hline \hline
         positions of particles ($N$ degrees of freedom) & $\leftrightarrow$ & parameters of the network ($N$ degrees of freedom) \\ \hline
         pairs of particles $(ij)$ & $\leftrightarrow$ & patterns $\mu$ \\ \hline
         energy ${\cal U}$ & $\leftrightarrow$ & loss ${\cal L}$ \\ \hline
         long range interaction & $\leftrightarrow$ & (for instance) cross-entropy \\ \hline
         finite range interaction & $\leftrightarrow$ & hinge loss \\ \hline
         particle density $\phi$ & $\leftrightarrow$ & number of data divided by the number of parameters $P/N$ \\ \hline
         separate two particles & $\leftrightarrow$ & fit a datum \\ \hline
         force distribution & $\leftrightarrow$ & density of unsatisfied patterns $P_+(\Delta)$ \\ \hline
         gap distribution & $\leftrightarrow$ & density of satisfied patterns $P_-(\Delta)$ \\ \hline
    \end{tabular}
    \caption{Correspondence between the jargon of particle systems and that of neural networks.}
    \label{tab:dict}
\end{table*}

\subsection{Deep Learning}
{\bf Set-up:} We consider a binary classification problem, with a set of $P$ distinct training data denoted as $\{\mathbf{x}_\mu,y_\mu\}_{\mu=1,\dots,P}$. The vector ${x}_\mu$ is the datum itself, which lives in dimension $d$ (e.g.\ it could be an image), and $y_\mu=\pm 1$ is its label. A network architecture corresponds to a function $f(\mathbf{x}; \mathbf{W})$, where $\mathbf{W}$ denotes the vector of parameters and $f(\mathbf{x};\mathbf{W})$  corresponds to the output of the network shown in Fig.~\ref{fig:network_architecture}.
In this scheme, each neuron sums the activity of all the neurons in the previous layer with some weights, sketched as connections in Fig.~\ref{fig:network_architecture} (each connection thus corresponds to one parameter $W^{(i)}_{\alpha,\beta}$). Next, a bias $B^{(i)}_\alpha$ is added to this sum (one additional parameter per neuron) to obtain the so-called pre-activation ($a^{(i)}_\alpha$ in the picture and in the equations). The neuron activity is then a non-linear function $\rho$ of that pre-activation: in what follows we will deal mainly with $\rho(a)=a\theta(a)$ --- the so-called rectified linear unit --- but we will also present some results with $\rho(a)=\mathrm{tanh}(a)$. The computation is done iteratively from the first layer (close to the input $\mathbf{x}$) to the last one (the output $f(\mathbf{x};\mathbf{W})$):
\begin{eqnarray}
    f(\mathbf{x};\mathbf{W}) \equiv a^{(L+1)},\vspace{0.5em}\\
    a^{(i)}_\beta = \sum_\alpha W^{(i)}_{\alpha,\beta}\,\rho\left(a^{(i-1)}_\alpha\right) - B^{(i)}_\beta,\\
    a^{(1)}_\beta = \sum_\alpha W^{(1)}_{\alpha,\beta}\,x_\alpha - B^{(1)}_\beta.
    \label{eq:recnnet}
\end{eqnarray}

\begin{figure}[ht]
    \centering
    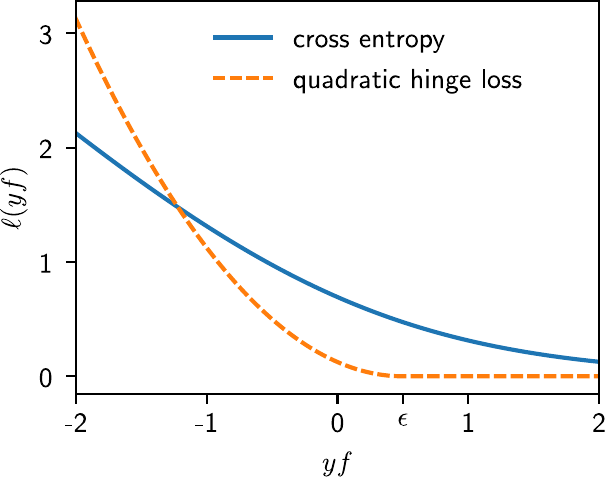
    \caption{Cross entropy and hinge loss functions. If the network classifies two classes with labels $y=\pm1$ then the loss can be written as $\ell(y, f) = \ell(yf)$. The plot shows the two cases studied in this work, namely the cross-entropy and the hinge loss; for the latter, a parameter $\epsilon = \frac12$ has been used.}
    \label{fig:loss_functions}
\end{figure}

In our notation the vector $\mathbf{W}$ contains all the parameters, including the biases. $\mathbf{W}$ is learned by minimizing a cost function, which can generically be written  $ \mathcal{L}(\mathbf{W}) = \frac1P \sum_{\mu=1}^P \ell\left(y_\mu, f(\mathbf{x}_\mu; \mathbf{W})\right)$. A widely chosen kind of loss is the cross entropy, $\ell(y,f) = \log\left(1+e^{-yf}\right)$. Another common choice is the hinge loss, defined as $\ell(y,f) = \frac12 \Delta(y,f)^2 \theta(\Delta(y,f)) = \frac12 \max(0, \Delta(y,f))^2$, where we have introduced the data overlap
\begin{equation}
    \Delta(y,f) \equiv \epsilon - y f,
\end{equation}
with $\epsilon>0$ being a constant. In what follows we choose $\epsilon=1/2$ without loss of generality \footnote{The parameter $\epsilon$ fixes the scale of the loss, in the sense that, if one rescales both $\epsilon$ and the weights of the last layer by the same factor $\alpha$, then all the observables are equivariant with respect to this transformation ($\Delta\to\alpha\Delta$, $\mathcal{L}\to\alpha^2\mathcal{L}$, $N_\Delta\to N_\Delta$, ...). Consequently, since any positive value of $\epsilon$ leads to the same behavior of the system, we have arbitrarily fixed its value to $\epsilon = \frac12$. If $\epsilon$ were $0$, the network would try to enforce $f(\mathbf{x};\mathbf{W})\equiv0$ regardless of the specific pattern.}. These loss functions take such a simple form only for a binary classification task, with labels $y=\pm1$, where $\ell(y,f) \equiv \ell(y\cdot f)$; the two loss functions are compared in Fig.~\ref{fig:loss_functions}. In the hinge loss, the condition $\Delta_\mu = \Delta\left(y_\mu, f(\mathbf{x}_\mu;\mathbf{W})\right)<0$ ensures that the datum $\mu$ is \textit{satisfied} --- that is, correctly classified by a margin $\epsilon$. The data which do not respect this margin will be referred to as \textit{unsatisfied} (not to be confused with \textit{misclassified} data, for which $y_\mu f(\mathbf{x}_\mu) < 0$) --- the number of such data will be denoted as $N_\Delta$. With this definition, $ \mathcal{L}$ is formally identical to ${\cal U}$ in Eq.~(\ref{1}) as already noted for the perceptron \cite{Franz15}, and it can be written as $ \mathcal{L}(\mathbf{W}) = \frac1P \sum_{\mu \in m} \frac12\Delta_\mu^2$, where $m$ is the set of unsatisfied patterns.  The correspondence between interacting particles and neural networks is summarized in Table~\ref{tab:dict}.

{\bf Performance of the hinge loss and its extension to multi-class problems:} This section can be skipped at a first reading.
We tested in the context of image classification that the hinge loss performs as well as the cross entropy on a state-of-the-art architecture~\cite{Gastaldi17}: we ran the implementation \footnote{\url{https://github.com/mariogeiger/pytorch_shake_shake}} for CIFAR-10 and we retrained it by replacing the cross entropy by the hinge loss. To compare the two losses in a standard setting we adapted the hinge loss for multiple classes, although in what follows we only study binary classification. To predict the label of an input $\mathbf{x}_\mu$ among $10$ possible labels $c=0,\dots,9$, the network's last layer returns as output a list of 10 values $f_{\mu,c}$: each $f_{\mu,c}$ can be interpreted as the probability that $c$ is the predicted label. Let $t_{\mu,c}$ be the true target labels: for each $\mu$, $t_{\mu,c}$ is equal to $1$ if $c$ is equal to the label of $\mathbf{x}_\mu$ and $-1$ otherwise. Multiclass hinge loss can then be written as
\begin{equation}
\mathcal{L} = \frac1{10P} \sum_{\mu,c} (\epsilon-t_{\mu,c}f_{\mu,c})^2 \theta(\epsilon - t_{\mu,c}f_{\mu,c}).
\end{equation}
We obtained an error of $3.72\%$ by running their original code (they report on github an error of $3.68\%$) and $3.61\%$, $3.65\%$, $3.82\%$ in three runs with the hinge loss.


\vspace{1em}

{\bf Effective number of parameters:} Following the argument developed after Eq.~(\ref{1}), we expect that at the transition point where the loss becomes non-zero, Eq.~(\ref{2}) will hold true and $N_\Delta\leq N$. (Related  arguments were recently made for a quadratic loss \cite{Cooper18}. In this case, we expect that the landscape will be related to that of floppy spring networks, whose spectra  were predicted in \cite{During13}). 
Just as is the case for the  jamming  of particles, here we must pay attention to the effective number of degrees of freedom that do affect the output, $N_\mathrm{eff}(\mathbf{W})$.  in the space of functions going from the neighborhoods of the training set to real numbers, we consider the manifold of functions $f(\mathbf{x};\mathbf{W})$ obtained by varying $\mathbf{W}$.  We denote by $N_\mathrm{eff}(\mathbf{W})$ the dimension of the tangent space of this manifold at $\mathbf{W}$. We discuss  in Appendix~\ref{app:neff} how $N_\mathrm{eff}(\mathbf{W})$ can be measured. In general we have $N_\mathrm{eff}(\mathbf{W})\leq N$. Several reasons can make $N_\mathrm{eff}(\mathbf{W})$ strictly smaller than  $N$, including:
\begin{itemize}
\item The signal does not propagate in the network, i.e. $f(\mathbf{x};\mathbf{W})=C_1$  where $C_1$ is a constant for all $\mathbf{x}$ in the neighborhood of the training points $\mathbf{x}_\mu$.
In that case, the manifold is of dimension unity and $N_\mathrm{eff}(\mathbf{W})=1$. This situation will occur for a poor initialization of the weights as discussed in~\cite{schoenholz2016deep}, or for example if all biases are too negative on the neurons of one layer  for ReLU activation function. It can also occur if the data $\mathbf{x}_\mu$ are chosen in an adversarial manner  for a given choice of initial weights. For example, one can choose input patterns so as to not activate the first layer of neurons (which is possible if the number of such neurons is not too large). Poor transmission will be enhanced (and adversarial choices of data will be made simpler) if the architecture presents some bottlenecks. In the situation where $N_\mathrm{eff}(\mathbf{W})=1$, it is very simple to obtain local minima of the loss at finite loss values, even when the model has many parameters.
\item The activation function is linear, then the output function is an affine function of the input, leading to $N_\mathrm{eff}\leq d+1$. Dimension-dependent bounds will also exist if the activation function is polynomial (because the output function then is also restricted to be polynomial).
\item Symmetries are present in the network, e.g. the scale symmetry in ReLU networks. It will reduce one degrees of freedom per node. 
\item Some neurons are never active e.g. in the ReLU case, their associated weights do not contribute to $N_\mathrm{eff}$.
\end{itemize}
Thus there are $N-N_\mathrm{eff}$ directions in parameters space that do not affect the function. These directions will lead to zero modes in the Hessian at any minima of the loss. In what follows we consider stability with respect to the $N_\mathrm{eff}$ directions orthogonal to those, which thus affect the output function. Our results on the impossibility to get stuck in bad minima are  expressed in terms of $N_\mathrm{eff}$. However, as reported in Appendix~\ref{app:neff}, we find empirically that for a proper initialization of the weights and rectangular fully connected networks, $N_\mathrm{eff}\approx N$ (the difference is small and equal to the number of hidden neurons, and only results from the symmetry associated with each ReLU neuron). Henceforth to simplify notations we will use the symbol $N$ to represent the number of effective parameters.
In the following sections, the Hessians are computed with respect to all the $N$ parameters.\vspace{1em}

{\bf Constraints on the stability of minima:} Let us suppose (and justify later) that for a fixed number of data $P$, if $N$ is sufficiently large then gradient descent with proper weights initialization leads to ${\cal L}=0$, whereas if $N$ is very small after training ${\cal L}>0$. Consider that $N$ is increased from a small value. At some value $N^*$ the loss obtained after training will approach zero \footnote{For finite $P$, $N^*$ will present fluctuations induced by differences of initial conditions. The fluctuations of $P/N^*$  are however expected to vanish in the limit where $P$ and $N^*$ become large. This phenomenon is well-known for the jamming of particles, and is referred to as finite size effects. }, i.e. $\lim_{N\rightarrow N^*}{\cal L}=0$. In analogy with the behavior of packings of particles, we refer to this point as the jamming transition. At the transition the stability constraint developed in Eq.~(\ref{4}) above also applies if the derivative of $f(\mathbf{x}; \mathbf{W})$ is continuous,  which holds true  if the non-linear function $\rho$ is smooth. Thus we have:
\be
\label{4bis}
P\geq N_\Delta \geq N_-\,,
\ee
since $P\geq N_\Delta$ (the number of unsatisfied patterns is obviously smaller than the total number of patterns). 

We shall assume  that  the fraction \mm{$N_-/N\equiv C_0$} of negative eigenvalues of ${\cal H}_p$ does not vanish in the large $N$ limit. In  Appendix~\ref{app:hpsym} we provide an argument supporting this result in the case of a specific non-linear function (ReLU) and random data, \mm{that yields $C_0=1/2$ independently of depth}. It implies that  unlike for spheres, but just like ellipses, ${\cal H}_p$ is not negative definite: we are therefore in the hypostatic scenario where one expects $N_\Delta<N^*$ at jamming, a point at which the spectrum must display a fraction of flat directions, as well as stiff ones, as described in Fig.~\ref{fig1}H. 

Moreover from this assumption and Eq.\ref{4bis}, we obtain that stability cannot be obtained for $N \geq P / C_0$.  For larger $N$, the dynamics cannot get stuck in a bad minimum, because in this over-parametrized regime there are not enough constraints to form them. It implies for the jamming transition that:
\be
\label{4ter}
N^*\leq P/C_0.
\ee
Notice that this bound is expected to be valid for any monotonic cost function, as for instance the cross entropy (the Hessian can always be decomposed as in Eq.~(\ref{3})). However, the spectrum of the Hessian would be different \footnote{For the cross entropy, when the data become separable true minima exists only  for diverging weights, a complication that does not occur with the hinge loss.}.\vspace{1em}

\begin{figure*}[ht!]
    \centering
    \setlength{\unitlength}{0.1\textwidth}
    \begin{picture}(9,3)
    \put(0,0){\def\svgwidth{0.29\textwidth}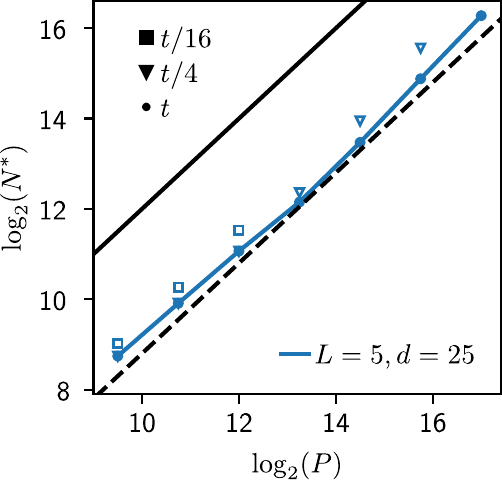}\put(2.5,2.9){(a)}
    \put(3,0){\def\svgwidth{0.29\textwidth}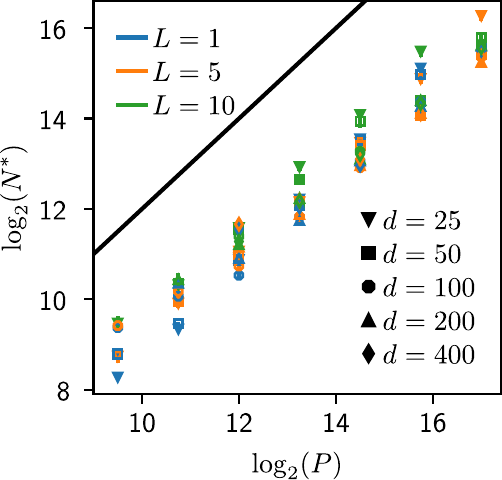}\put(5.5,2.9){(b)}
    \put(6,0){\def\svgwidth{0.29\textwidth}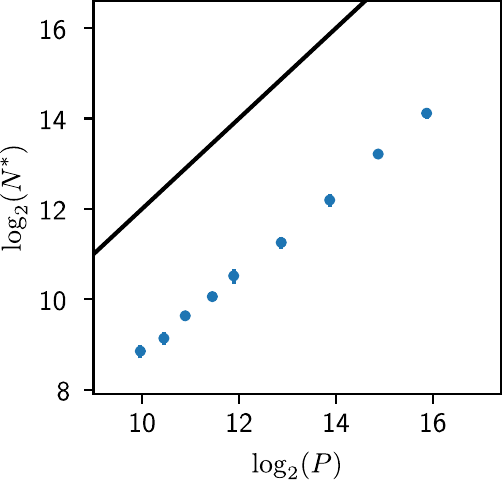}\put(8.5,2.9){(c)}
    \end{picture}
    \caption{Jamming transition with random data. (a) $N^*$ vs number of data $P$ for different learning times as indicated in legend, where $t=10^6$ steps and a cross-entropy loss function is used. The curves at small times (orange and green) are shown as diverging to indicate the absence of the transition. The dotted black line toward which the dynamics appear to converge has slope 1, supporting $N^*\sim P$ at long times. Here $L=5$ and $d=25$. (b) $N^*$ vs number of data $P$ after $t=10^6$ for various depths $L$ and input dimensions $d$ as indicated in legend, using the same loss function. The transition shows little dependency on $L$ and $d$. (c) Same plot as (b) for a network with hinge loss, with $d=h$ and $t=2\cdot 10^6$.
    In the three plots (a,b,c), the black line indicates the theoretical upper bound: $P/N^*=1/2-N_c/N$ derived for the hinge loss.
    }
    \label{fig:transition}
\end{figure*}
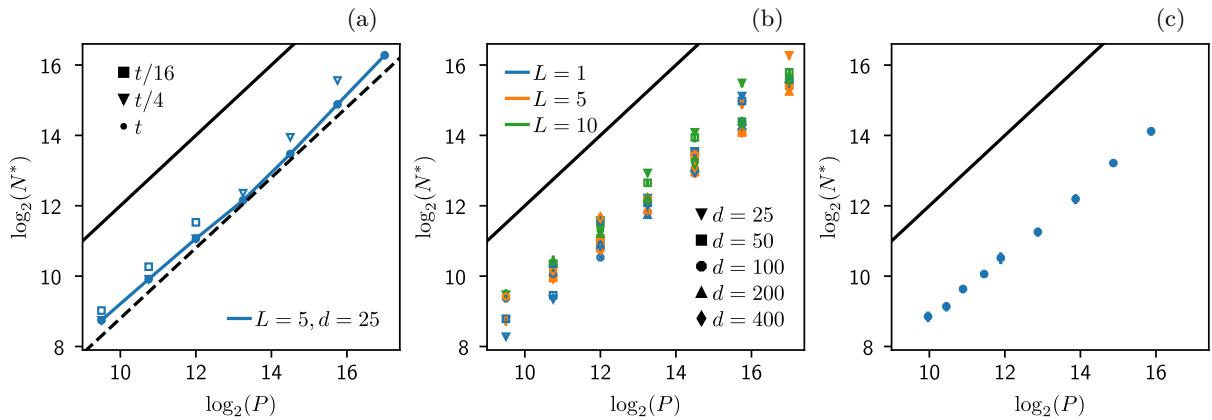

{\bf Smooth {\it vs} non-smooth output function:} In our numerical study below, we consider the most common choice for the non-linear function $\rho$, namely the rectified linear unit (ReLU): $\rho(a) = a\,\Theta(a) = \max(0, a)$. In that case, as stated above we expect for random data the spectrum  of ${\cal H}_p$ to be symmetric (a fact that appears to also hold true for the image dataset we use, see below), thus $N_-/N=1/2$. Yet, with the ReLU, $f(\mathbf{x}; \mathbf{W})$ is not continuous and presents cusps,  so that the Hessian needs not be positive definite for stability and Eq.~(\ref{4bis}) needs to be modified. Introducing the number of directions $N_c$ presenting cusps, stability implies  $N_\Delta\geq N_--N_c$ leading to $C_0 N^*\leq P+N_c$.
Empirically  we find that  $N_c/N^*\in [0.21, 0.25]$ both for random data and images as reported in Appendix \ref{app:zeros}, implying that:
\be
N^*\leq 4P.
\ee

For comparison, below we also present results for networks with $\mathrm{tanh}$ activation functions. In that case the landscape is smooth and the system ends up in  minima without negative eigenvalues. For such networks the spectrum of $\mathcal{H}_p$ is not exactly symmetric, and we observe $C_0 = N_-/N \approx 0.43$.

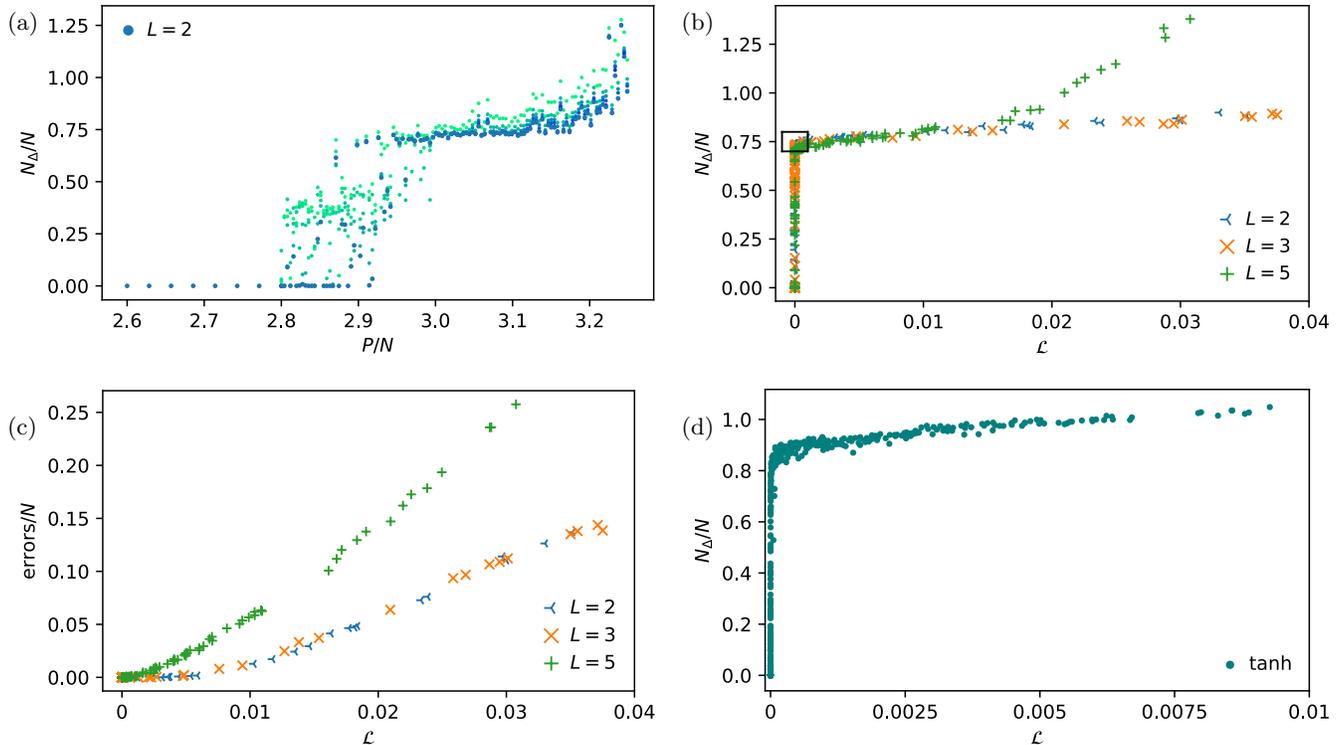
\begin{figure*}[ht]
    \centering
    \setlength{\unitlength}{0.1\textwidth}
    \begin{picture}(10,5.9)
    \put(0.1,2.9){\def\svgwidth{0.47\textwidth}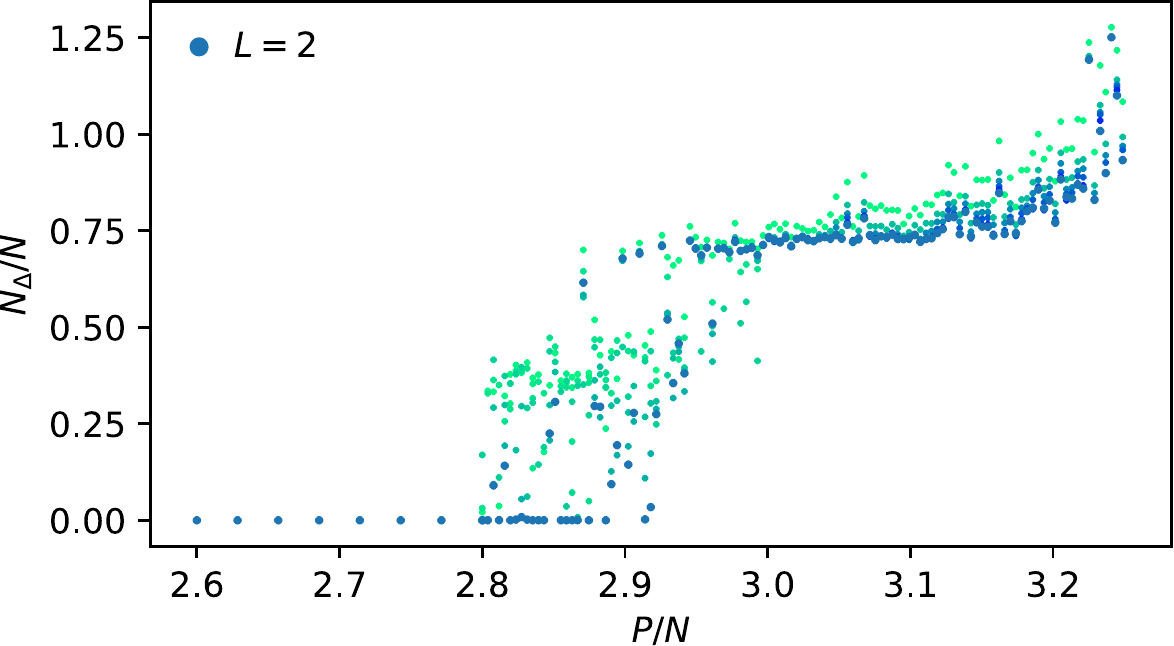}\put(0,5.3){(a)}
    \put(5.1,2.9){\def\svgwidth{0.47\textwidth}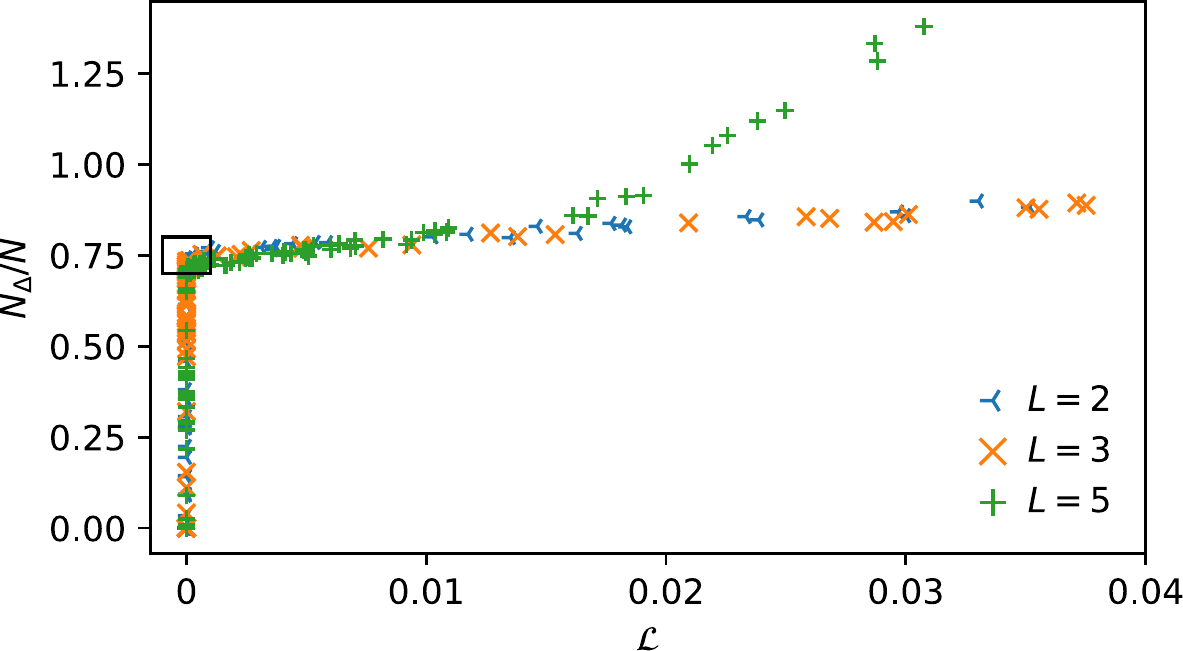}\put(5,5.3){(b)}
    \put(0.1,0){\def\svgwidth{0.47\textwidth}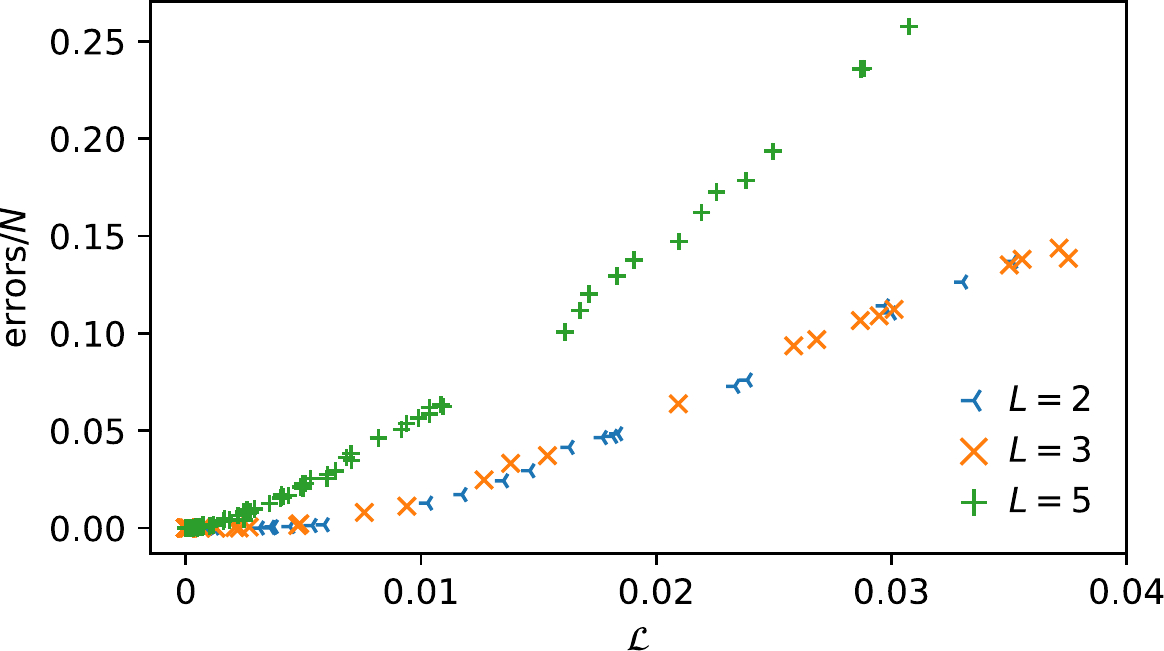}\put(0,2.3){(c)}
    \put(5.1,0){\def\svgwidth{0.47\textwidth}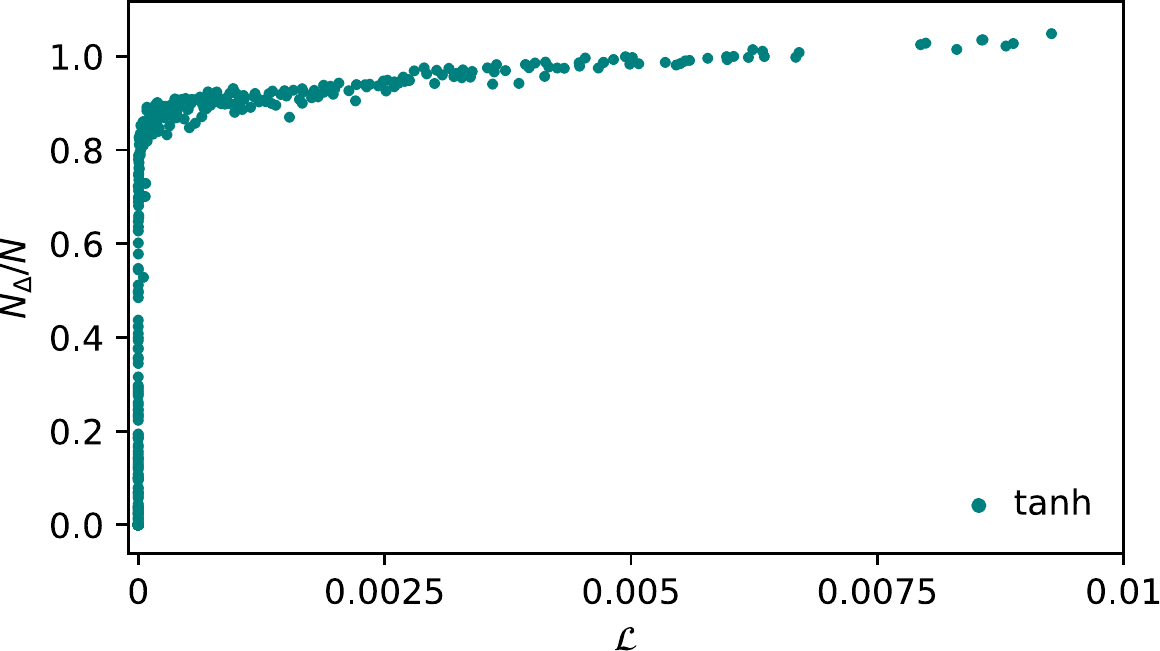}\put(5,2.3){(d)}
    \end{picture}
    \caption{Behaviour near the transition for random data. 
    (a) Number of unsatisfied constraints $N_\Delta$ per parameter $N$ as a function of $P/N$. Collections of vertical points correspond to the same run, but with different learning times from green (short time, starting at $3\cdot10^5$ steps) to blue ($10^7$ steps). The data support a discontinuous jump in this quantity at some $P/N\in [2.8,2.9]$ at asymptotically long times. Indeed, outside that range the learning dynamics appear to have converged to zero for $r<2.8$, and to some value $> 0.7$ for $P/N>2.9$. In the interval $P/N\in [2.8,2.9]$, data are still evolving in time. (b) $N_\Delta/N$ vs $\mathcal{L}$ follows a curve with almost no scatter for all $L=2,3,5$. This is similar to the jamming transition where finite size noise is eliminated when quantities are plotted against the potential energy, rather than the packing fraction~\cite{Liu10}. The black rectangle on the left side of the plot (small loss $\mathcal{L}$ and finite ratio $N_\Delta/N$) marks the points in the underparametrized phase that are close to the transition. (c) Relationship between the number of misclassified data (data points with negative $y_\mu f(\mathbf{x}_\mu)$) and $\mathcal{L}$, displaying a smooth behavior. (d) $N_\Delta/N$ vs $\mathcal{L}$ for a network with $L=3$, but with tanh activation functions rather than ReLUs.}
    \label{fig:errorloss}
\end{figure*}

{\bf Main results:} Overall, our analysis supports that
\begin{enumerate}
    \item In the case of hinge loss there is a sharp transition for $N^*\leq P/C_0$ ($N^*<4P$ with the ReLU), below which  the loss converges to some non-zero value (under-parametrized phase) and above which it becomes null (over-parametrized phase).
    \item At that point the fraction $N_\Delta/N$ of unsatisfied constraints per degree of freedom jumps to a finite value, see Fig.~\ref{fig:errorloss} (a,b). 
    \item Unlike for spheres or the perceptron, isostaticity $N_\Delta/N=1$ cannot be guaranteed. Instead one expects generically $N_\Delta/N<1$ as for ellipses.
    \item We are thus in the hypostatic universality class, where the scaling properties of the spectrum of the Hessian near jamming are prescribed in Fig.\ref{fig1}.
\end{enumerate}
In the  next sections, we confirm these predictions in numerical experiments and observe the generalization properties at and beyond the transition point.

\section{For random data the transition  occurs for \texorpdfstring{$\boldsymbol{N\sim P}$}{N vs P}}\label{satunsat}

We begin the numerical study of the transition between the overparametrized and underparametrized regime in the case of random data, taken to lie on the $d$-dimensional hyper-sphere of radius $\sqrt{d}$, ${\bf x}_\mu\in {\cal S}^d$ with random label $y_\mu=\pm 1$. The source code used to generate the simulations described in this section and the following ones is available at \url{https://github.com/mariogeiger/nn_jamming}. We proceed as follows: we build a network with a number of weights $N$ large enough for it to be able to fit the whole dataset without errors. Next, we reduce the number of weights by decreasing the width $h$ while keeping the depth $L$ fixed, until the network cannot correctly classify all the data anymore within the chosen learning time. We denote this transition point $N^*$.

We have noticed (data not shown) that the precise location of the transition point $P/N^*$  has a mild dependence on the dynamics (ADAM versus regular SGD, choice of batch size, learning rate schedule, etc\ldots): the same holds true for the jamming of repulsive particles, where the choice of the dynamics affects the precise value of the critical density $\phi_c$, but not the critical behaviour close to this point.

{\bf Cross-entropy loss:} We first consider the cross-entropy loss --- the results are qualitatively similar to those with the hinge loss. As initial condition for the dynamics we use the default initialization of {\tt pytorch} \footnote{Weights and biases are initialized with a uniform distribution on $[-\sigma, \sigma]$, where $\sigma^2 = 1/f_{in}$ and $f_{in}$ is the number of incoming connections.}. 
The system then evolves according to a stochastic gradient descent (SGD) with a learning rate of $10^{-2}$ for $5\cdot10^5$ steps and $10^{-3}$ for $5\cdot10^5$ steps; the batch size is set to $\min(P/2, 1024)$; only in this case, with the cross-entropy loss, batch normalization is also used. In Fig.~\ref{fig:transition} (a) we show how $N^*$ depends on the total learning time: the larger is the learning time the more the asymptotic relationship $N^*$ vs $P$ is consistent with an asymptotic linear behaviour. Note that for large $P$ and small times, errors are always present and the transition cannot be found. 

In Fig.~\ref{fig:transition} (b) we show $N^*$ versus the number of data $P$ after $t=10^6$ steps for several depths $L$ and input dimensions $d$ (we checked that $t=10^6$ is enough to get convergence to the conjectured asymptotic linear behaviour for all depths investigated). It is noteworthy that (i) the points always lie below the theoretical upper bound $P/N^*=1/2-N_c/N$, and (ii) the transition does not appear to depend on $L$ and $d$. Surprisingly, this result indicates that in the present setup the ability of fully connected networks to fit random data is independent of the depth. As we shall see, \mm{we observe the same independence on depth} for the image data studied below.

{\bf Hinge loss:} In order to test the dependence of our results on the specific choice of the loss function, we performed the same experiment using the hinge loss. In this case we used an orthogonal initialization~\cite{Saxe13}, no batch normalization and $t=2\cdot10^6$ steps of ADAM~\cite{Kingma14} with batch size $=P$ and a learning rate starting at $10^{-4}$, progressively divided by $10$ every 250k steps. The location of the transition is shown in Fig.~\ref{fig:transition} (c): results are very similar to that of the cross-entropy loss.

{\bf Hinge {\it v.s.}  cross-entropy loss from a conceptual perspective:} As shown above, both losses appears to lead to similar performances. As shown in this section, both of them also displays a transition where all data are fitted. Yet, the nature of this transition is harder to investigate for the cross-entropy. Indeed in that case the total loss is never zero, except if the output and therefore the weights diverge. Thus in the over-parametrized phase, the learning dynamic never settles, and the weights slowly drift to infinity. In practice, users stop learning at finite times (which is not needed for the hinge loss where the dynamics really stops in the over-parametrized regime when the loss vanishes). Working at finite time however blurs true critical behavior near jamming, as discussed in \cite{geiger2019scaling}.

\section{The transition is hypostatic}

From the analysis of Section~\ref{analogy}, the number of constraints per parameter $N_\Delta/N$ is expected to jump discontinuously at the transition. To test this prediction we consider several architectures, both with $N \approx 8000$ and $d=h$ but with different depths $L=2$, $L=3$ and $L=5$. The vicinity of the transition is studied by varying $P$ around the transition value. We used the hinge loss with the same gradient descent dynamics as described above, for a duration of $10^7$ steps.
Fig.~\ref{fig:errorloss} (a) reports the ratio $N_\Delta/N$ as a function of the ratio $P/N$ and of the learning time, as detailed in caption. It is clear that in the range where $N_\Delta/N$ has reached a stationary value (i.e. for $P/N<2.8$ and $P/N>2.9$), a jump has occurred from 0 to $N_\Delta/N\approx 0.75$, a result consistent with the bound of  Eq.~(\ref{4}) implying $N_\Delta/N\geq (N_- - N_c)/N \gtrapprox 0.25$. For $P/N\in [2.8,2.9]$, the dynamics has not yet converged and the data are somewhat scattered. This observation is presumably the signature of the usual slowing down that occurs near critical points.

Fig.~\ref{fig:errorloss} (b) shows the same quantity $N_\Delta/N$, now plotted as a function of the loss $\mathcal{L}$. Strikingly, all the scatter is gone, and one observes a clear discontinuous behaviour for ${\cal L}\rightarrow 0$. Interestingly, this state of affairs is very similar to the jamming transition of particles, for which the noise in the data due to finite size effects is quite strong when quantities are expressed in terms of the density $\phi$ (analogous to $P/N$) but very small when quantities are expressed in terms of potential energy ${\cal U}$ (analogous to ${\cal L}$)~\cite{Ohern03}.

For the sake of completeness we also show the number of misclassified data as a function of the loss in Fig.~\ref{fig:errorloss} (c). The number of misclassified data increases monotonically --- and initially very slowly ---  with the loss. Indeed, close to the jamming threshold in the underparametrized phase, if $0<\Delta_\mu<\epsilon$ the pattern $\mu$ is well classified but the corresponding gap $\Delta_\mu$ is positive: unsatisfied constraints do not lead to misclassification right away.

In Fig.\ref{fig:errorloss} (d) we show that $N_\Delta/N$ vs the loss $\mathcal{L}$ exhibits a sharp transition also for networks with tanh activation functions.

\section{Spectrum of the Hessian of the loss  near Jamming }\label{spectrum}

\begin{figure*}[ht]
    \centering
    \setlength{\unitlength}{0.1\textwidth}
    \begin{picture}(10,4.5)
    \put(-0.7,2.8){\def\svgwidth{0.28\textwidth}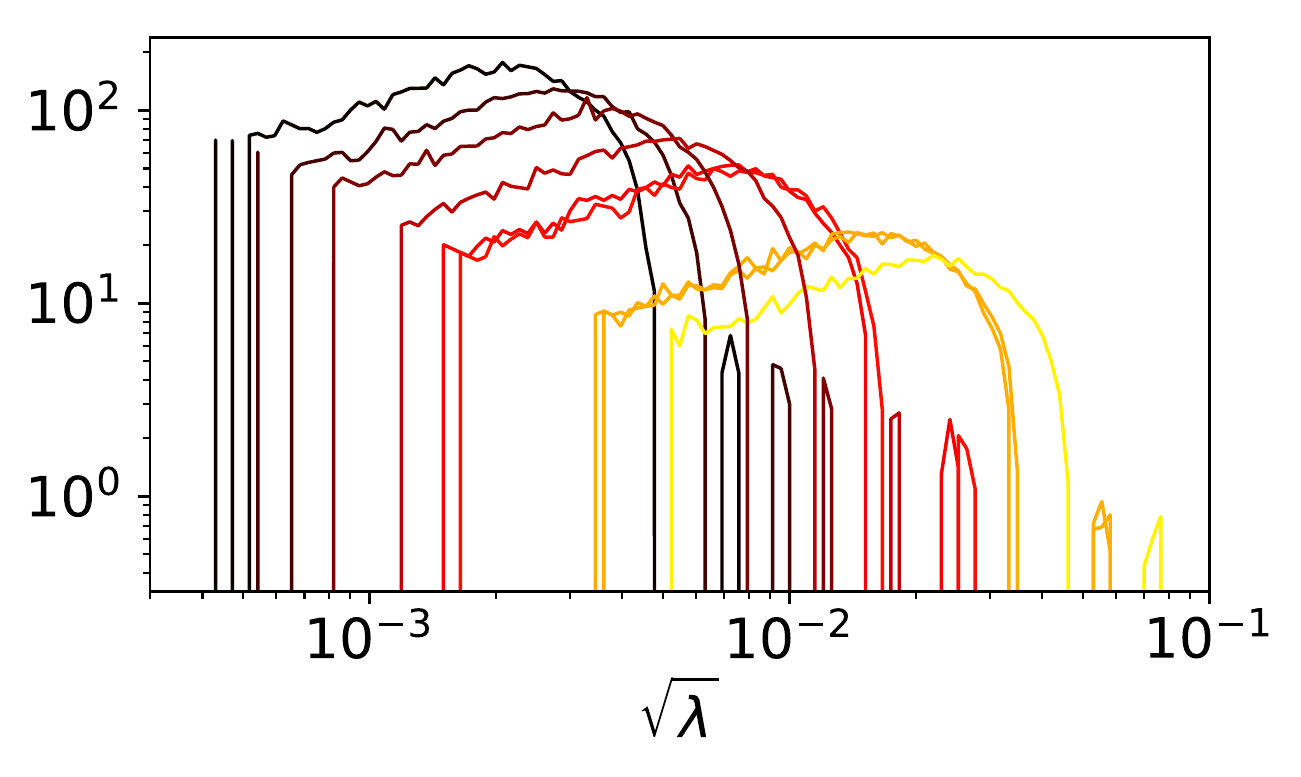}\put(0.35,4.5){(a) $\quad\mathcal{H}_p$}
    \put(2.05,2.78){\def\svgwidth{0.28\textwidth}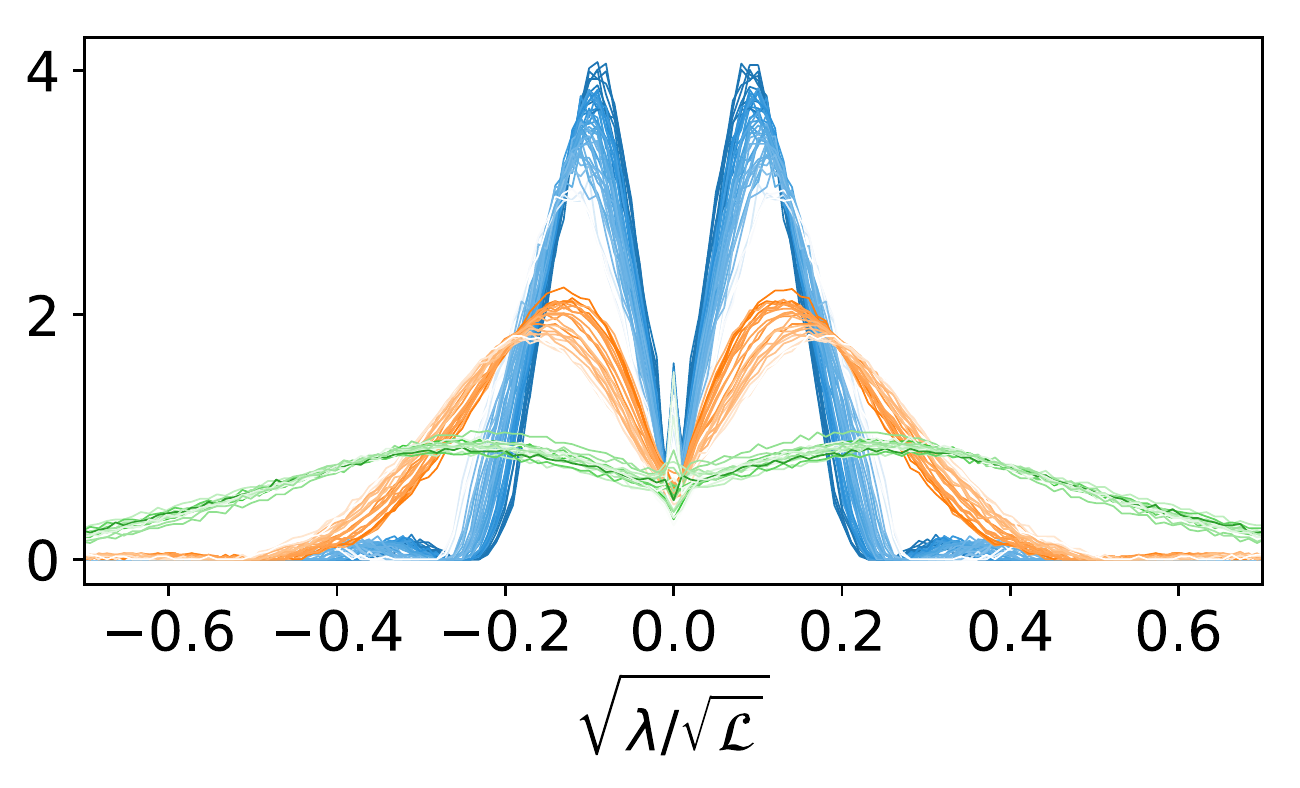}\put(3.05,4.5){(b) $\quad\mathcal{H}_p$}
    \put(4.8,2.8){\def\svgwidth{0.28\textwidth}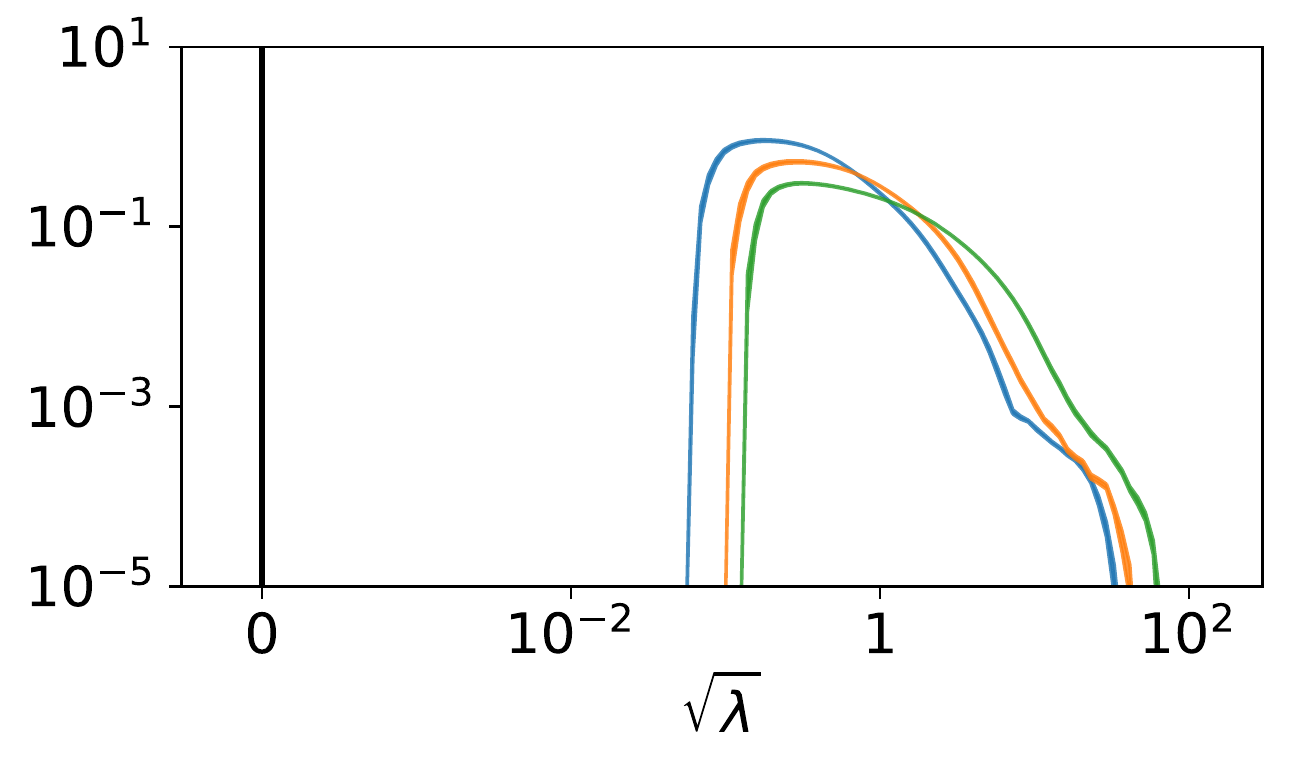}\put(5.85,4.5){(c) $\quad\mathcal{H}_0$}
    \put(7.6,2.78){\def\svgwidth{0.28\textwidth}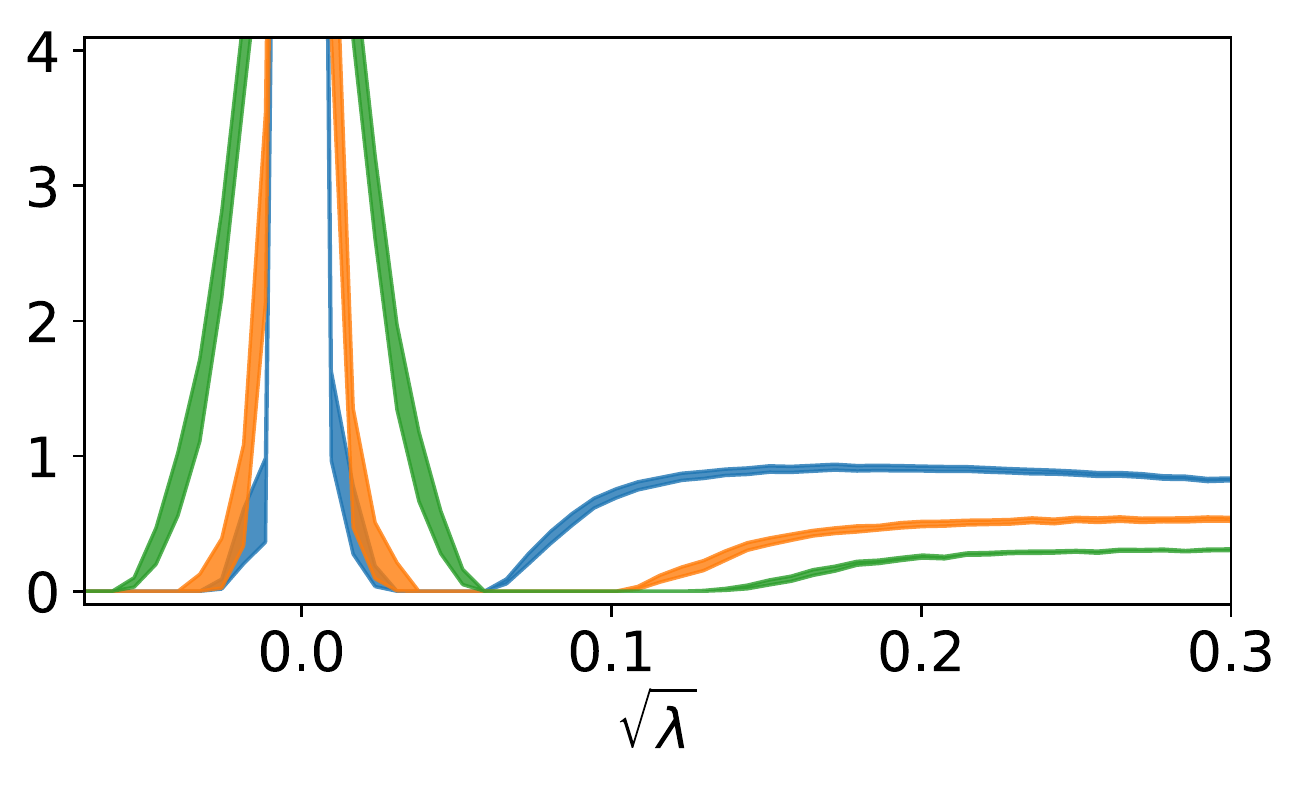}\put(8.65,4.5){(d) $\quad\mathcal{H}$}
    \put(0.1,0){\def\svgwidth{0.3\textwidth}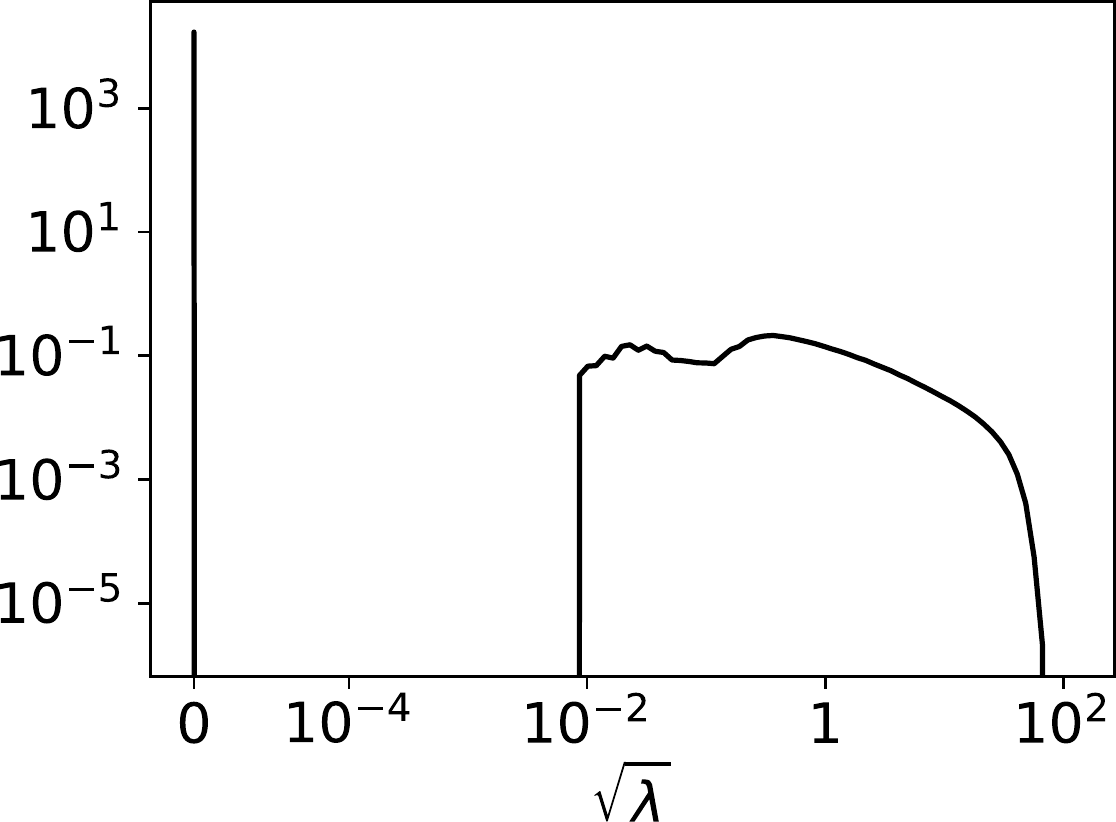}\put(1.35,2.35){(e) $\quad\mathcal{H}_p$}
    \put(3.4,0.05){\def\svgwidth{0.3\textwidth}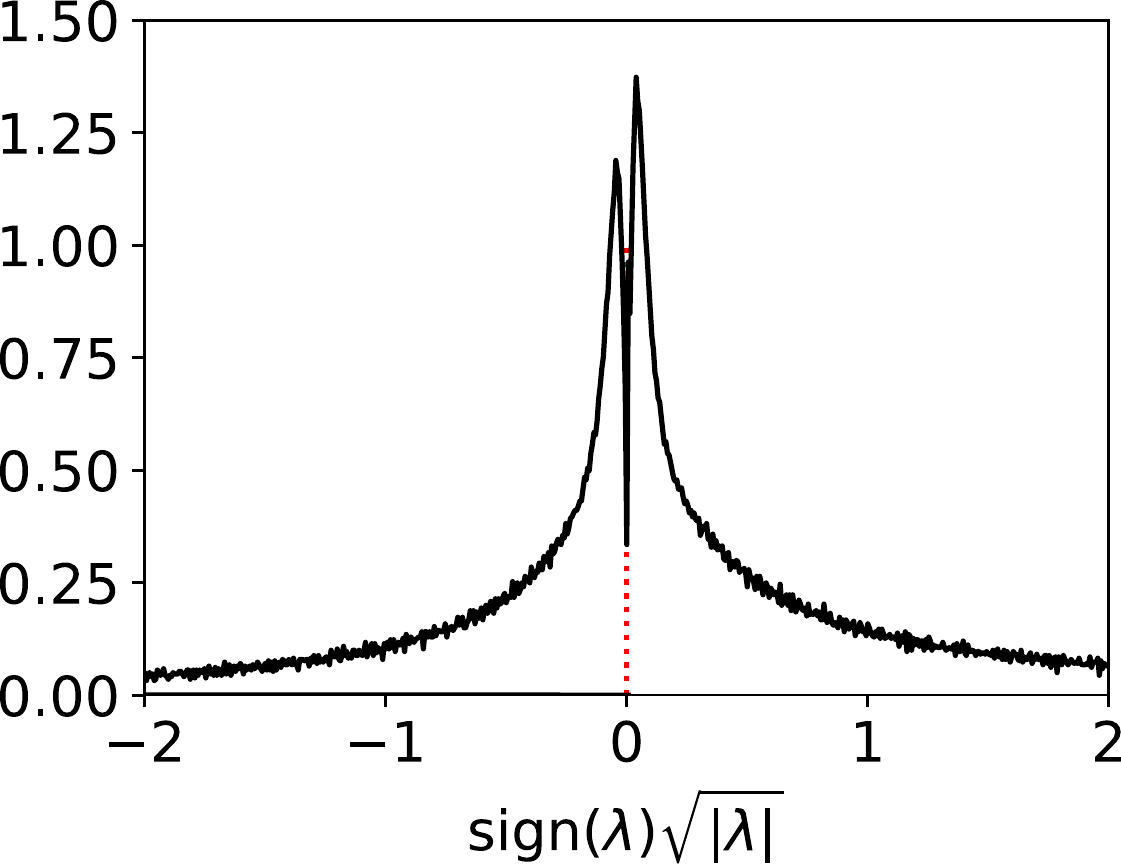}\put(4.7,2.35){(f) $\quad\mathcal{H}_0$}
    \put(6.7,0){\def\svgwidth{0.3\textwidth}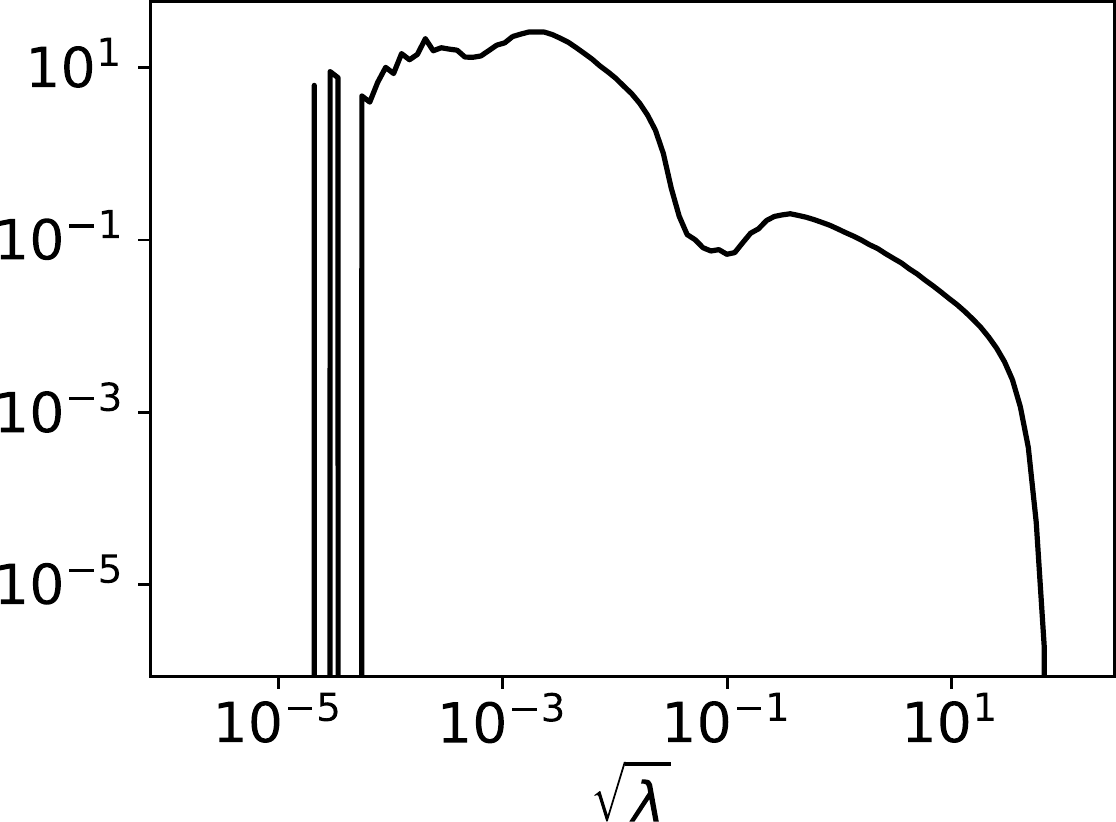}\put(8.05,2.35){(g) $\quad\mathcal{H}$}
    \end{picture}
    \caption{The data shown in this figure concern the underparametrized points close to the transition for random data, which in Fig.~\ref{fig:errorloss} (b) are enclosed in a black rectangle. (a) Positive part of the spectrum of ${\cal H}_p$ for  ten distinct runs  in the underparametrized phase close to the transition. The associated loss value grows from black (low) to yellow (high). (b) These spectra collapse when plotted in terms of $\lambda/\sqrt{\mathcal{L}}$ as expected. Lighter colors correspond to higher losses. Note that they appear symmetric, in agreement with our hypothesis estimating the number of negative modes (an argument that explains this fact can be found in Appendix~\ref{app:hpsym}). Colors are as in (d): $L=2$ (blue), $L=3$ (red) and $L=5$ (green). (c) The spectrum of ${\cal H}_0$  contains a delta function in zero of weight $N-N_\Delta$, followed by a gap, followed by a continuous spectrum, as expected for hypostatic systems. (d)  The spectrum of the total Hessian $\cal H$ has a similar shape, excepted that the delta function is blurred. Note that $\cal H$ has negative eigenvalues. These directions may in fact be stabilized by the $N_c$ cusps of the linear rectifier, or alternatively may indicate  that the learning dynamics did not converge to a local minimum yet. The thickness of each line correspond to the standard deviation. 
    (e-g) Spectrum of the matrices $\mathcal{H}_p, \mathcal{H}_0$ and $\mathcal{H}=\mathcal{H}_0+\mathcal{H}_p$ for tanh networks, respectively. Notice that the spectrum of $\mathcal{H}_p$ is no longer symmetric, compared to the ReLU case: the faction of negative eigenvalues is $C_0\approx0.43$. \label{fig:hessian}}
\end{figure*}

The Hessian is a key feature of landscapes, as it characterizes its curvature, and it is also a central aspect of the theoretical description above. In this section we systematically analyze the spectra of $\mathcal{H}$, $\mathcal{H}_0$ and $\mathcal{H}_p$. To test the predictions on the singularity of the Hessian matrix, we need to focus on the underparametrized data points near the transition. These points are contained in the black rectangle on the left side of Fig.~\ref{fig:errorloss} (b). The networks that we use are relatively small, but, for reference, it would be possible to compute the spectrum of the Hessian also for large networks, as discussed e.g.~in~\cite{adams2018estimating,ghorbani2019investigation}. The setting is as above: the network uses the hinge loss and is trained with ADAM with full batches (batch size $=P$), orthogonal initialization and no batch normalization.

{\bf Relu networks:} At the end of each run, we compute the hessian $\cal H$ of the loss $\cal L$, as well as the two terms ${\cal H}_0$ and ${\cal H}_p$ contributing to it, as defined in Eq.~(\ref{3}).
Fig.~\ref{fig:hessian} (a) shows the positive part of the spectrum of ${\cal H}_p$ for different values of the loss, illustrating that the dependence on the latter is very significant. In Fig.~\ref{fig:hessian} (b) we confirm that the spectrum of ${\cal H}_p$ collapses when the eigenvalues are re-scaled by ${\cal L}^{1/2}$, as expected from Section~\ref{analogy}. The key observation is that these spectra are symmetric, as argued in Appendix~\ref{app:hpsym}. We also don't observe any accumulation of eigenvalues at $\lambda=0$, except for the trivial zero modes stemming from the scaling symmetry of ReLU neurons (whose number is the total number of hidden neurons, much smaller than the number of weights). Fig.~\ref{fig:hessian} (c) shows the spectrum of $\mathcal{H}_0$ at the end of training for runs close to the jamming transition. As expected it is semi-positive definite, with a delta peak at $\lambda=0$ corresponding to $N-N_\Delta$ modes. It is followed by a gap and a continuous spectrum, as predicted near the jamming transition of particles if $N_\Delta<N$ \cite{During13} (which occurs for elliptic particles \cite{Brito18a}). As the loss increases, $N_\Delta$ increases and the gap is reduced. Finally in Fig.~\ref{fig:hessian} (d), the spectrum of $\cal H$  is shown. Interestingly the spectrum of the Hessian is not positive definite, but present some unstable modes. This phenomenon stems from our choice of ReLU activation function, which leads to cusps in the landscape as quantified in the~\ref{app:zeros}. Such cusps can stabilize directions that would be unstable according to the Hessian.

{\bf Tanh networks:} On the contrary, networks with $\mathrm{tanh}$ activation functions exhibit a smooth landscape, and in principle the loss is able to reach minima without any negative eigenvalues, since there are no cusps that could possibly stabilize them. Indeed, when minimizing a $\mathrm{tanh}$-network with $P=11000$ random patterns and $N=2232$ parameters, we observe that after 10 million ADAM steps there remain only 11 negative eigenvalues (between $-5\cdot10^{-5}$ and $-2\cdot10^{-8}$), and after 100 million ADAM steps only 6 were left (between $-4\cdot10^{-6}$ and $-2\cdot10^{-8}$). For comparison, in ReLU networks the number of negative eigenvalues at the end of training is about $10\%N$.

In Fig.~\ref{fig:hessian} (e-g) we show the spectrum of the matrices $\mathcal{H}_0, \mathcal{H}_p$ and $\mathcal{H}=\mathcal{H}_0+\mathcal{H}_p$, for a $\mathrm{tanh}$-network at jamming. The matrix $\mathcal{H}_0$ is qualitatively similar to what was observed in ReLU-networks: it presents a delta peak in $0$ and a gapped bulk of positive eigenvalues. The matrix $\mathcal{H}_p$ appears quite different, since it is no longer symmetric: the number $N_-$ of negative eigenvalues is approximately $0.43 N$ --- therefore $C_0=0.43$ instead of $C_0=0.5$.
The total Hessian $\mathcal{H}$ is not gapped in the present case, even though it displays two peaks. In order to clearly have a gap we would have to sample points closer to jamming (with a smaller loss, since $\mathcal{H}_p$ is proportional to $\sqrt{\mathcal{L}}$ close to jamming).

Overall, as we move from the under-parametrized phase to the over-parametrized one the situation is as follows:
\begin{enumerate}
    \item $N$ below $N^*$: There are many constraints with respect to the number of variable $N$, $\mathcal{H}_0$ is almost full rank and can easily compensate the negative eigenvalues of $\mathcal{H}_p$. The spectrum of $\mathcal{H}_p$ is symmetric.
    \item $N$ approaching $N^*$ from below: The rank of $\mathcal{H}_0$ decreases but it does not go below $C_0 N$ since it has to compensate the vanishingly small negative eigenvalues of $\mathcal{H}_p$.
    \item As $N$ is large enough, the dynamics finds a global minimum at $\mathcal{L}=0$ and $\mathcal{H}_p$ vanishes.
\end{enumerate}

\section{Distribution of gaps reveals new singular behaviour}

\begin{figure*}[ht]
    \centering
    \setlength{\unitlength}{0.1\textwidth}
    \begin{picture}(10,6)
    \put(0.2,3){\def\svgwidth{0.47\textwidth}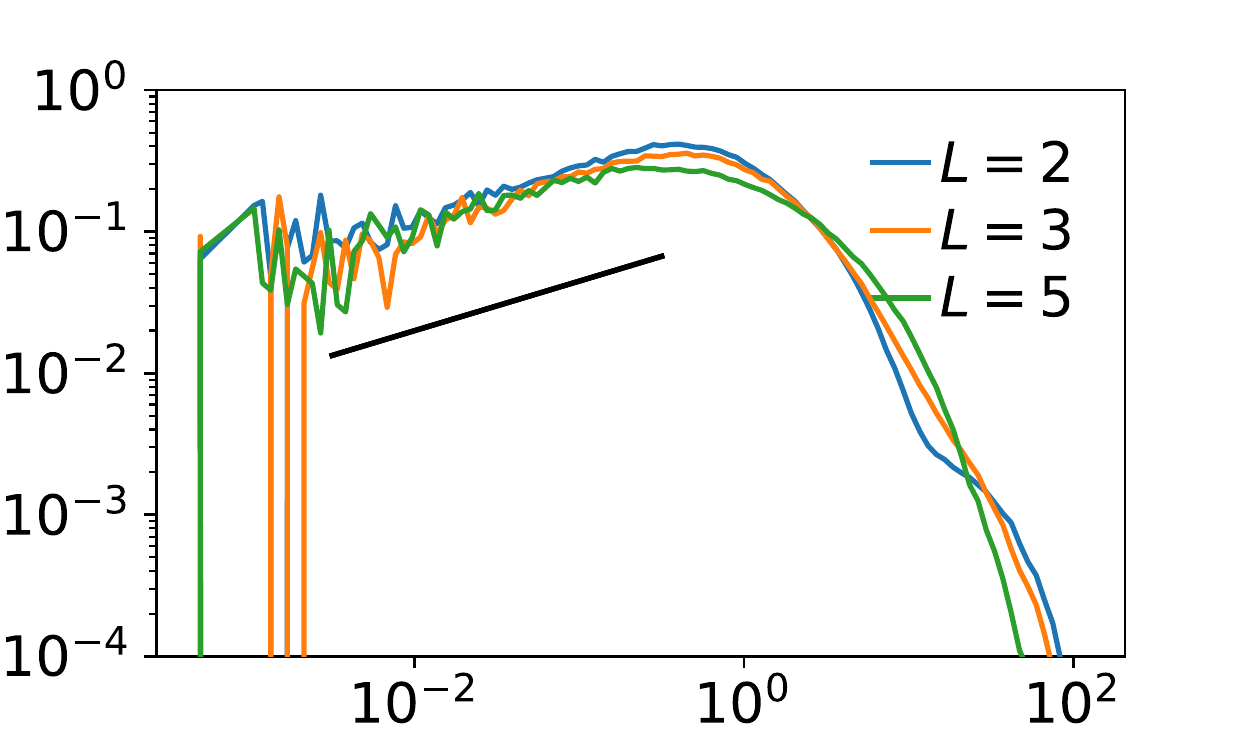}\put(0,5.6){(a)}
    \put(5.1,3){\def\svgwidth{0.47\textwidth}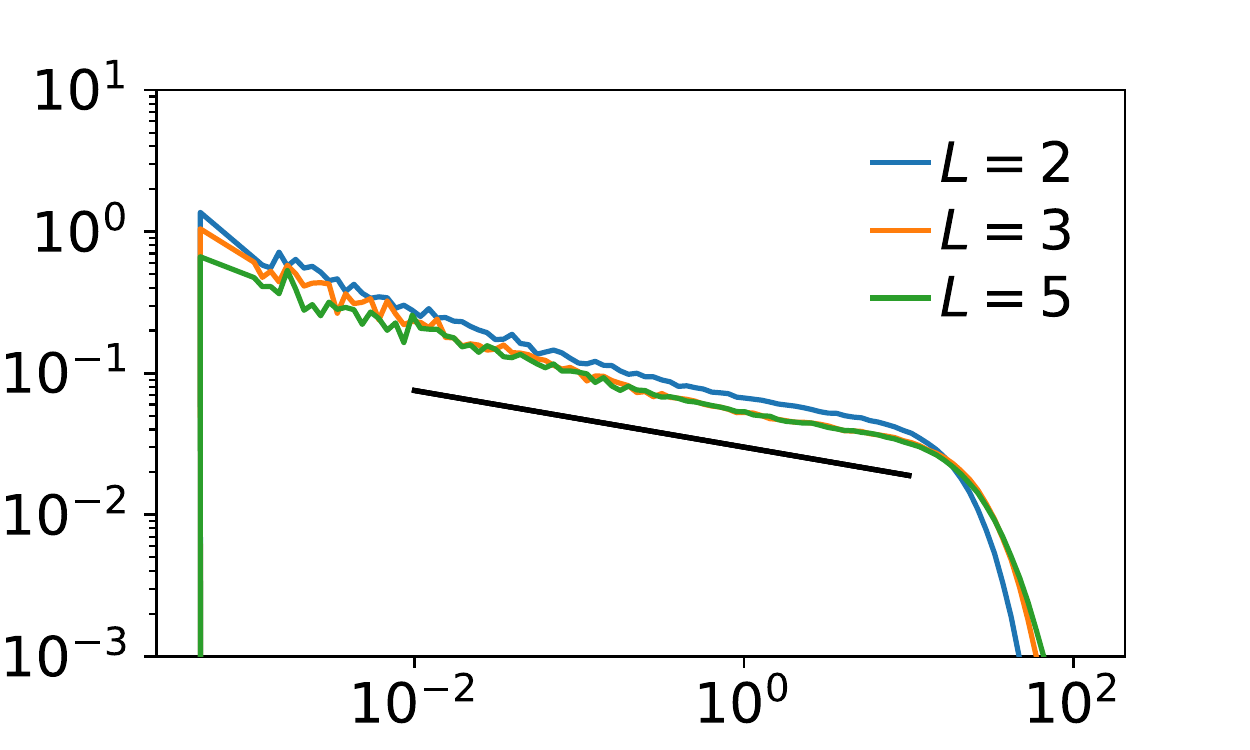}\put(4.9,5.6){(b)}
    \put(0.2,0.1){\def\svgwidth{0.47\textwidth}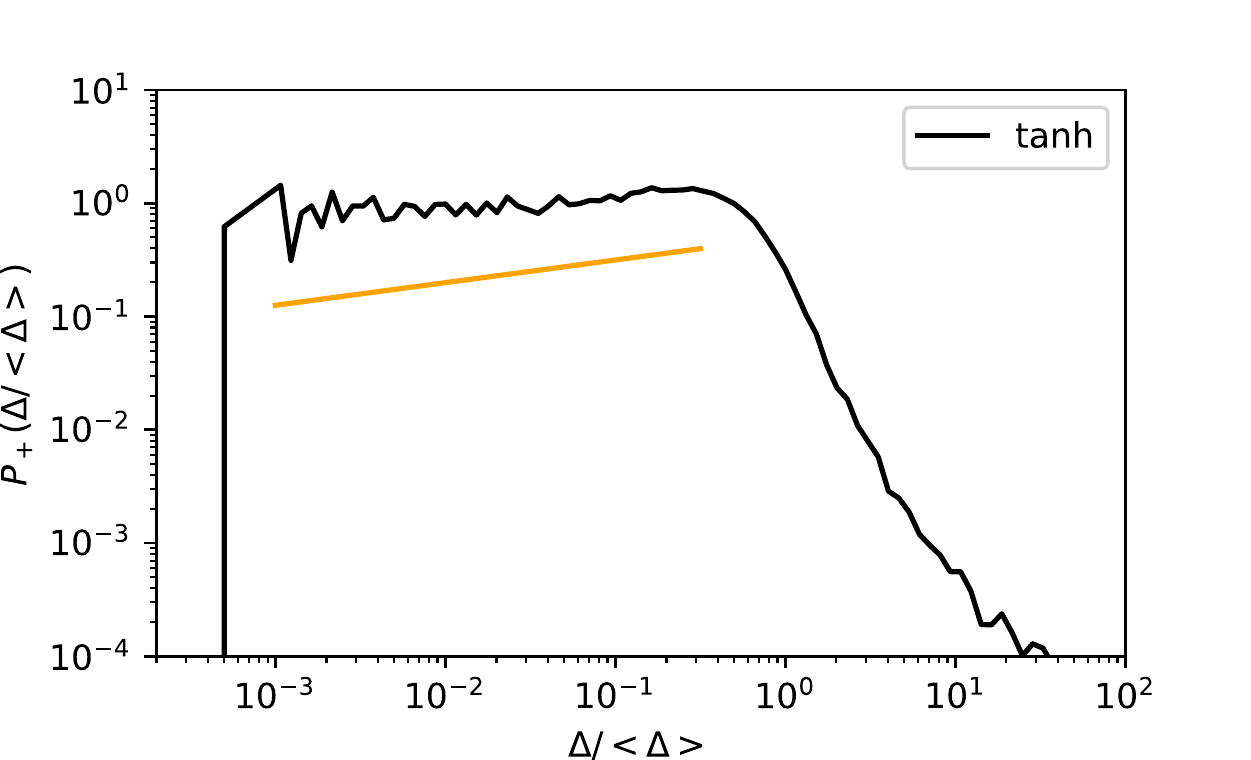}\put(0.1,2.6){(c)}
    \put(5.1,0.1){\def\svgwidth{0.47\textwidth}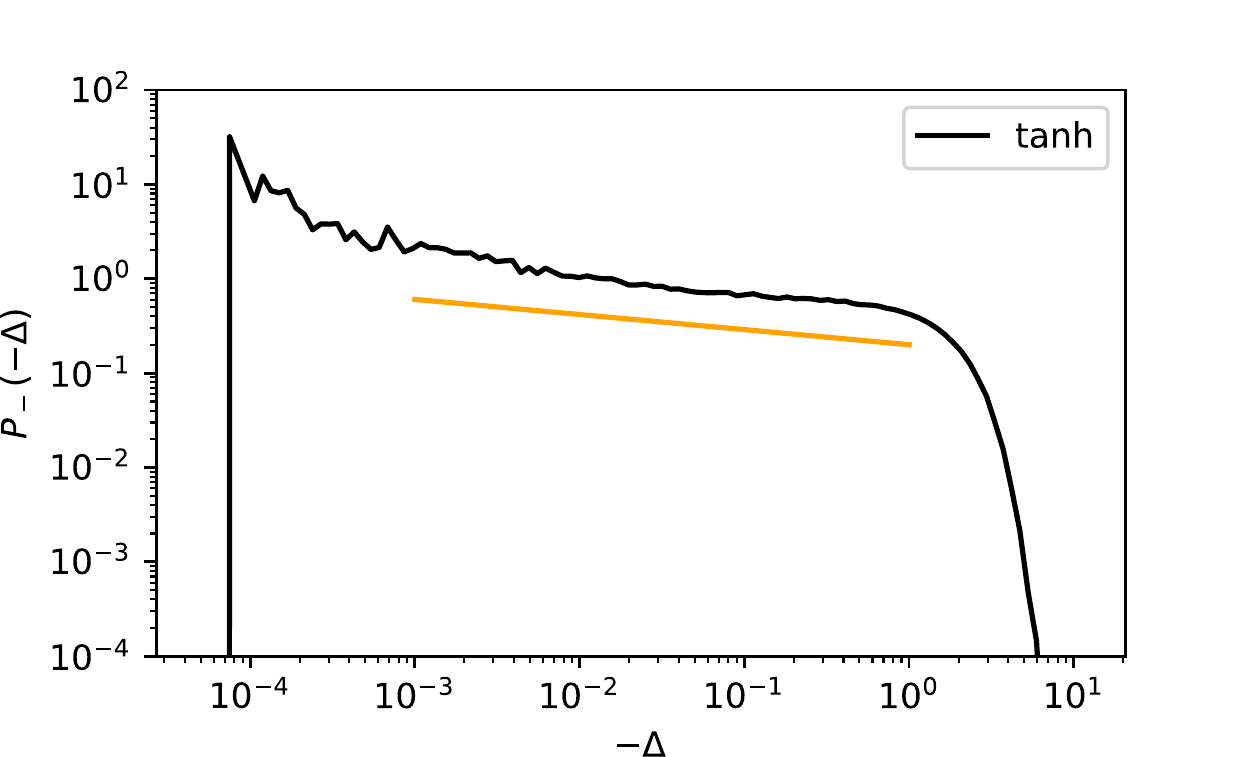}\put(4.9,2.6){(d)}
    \end{picture}
    \caption{(a) Distribution of re-scaled overlaps $z\equiv\Delta/\sqrt{\mathcal{L}}>0$ near threshold, supporting that $P_+(z)\sim z^\theta$ with an exponent $\theta\approx 0.3$ that does not vary with $L$ in the range probed. 
    (b) The distribution of gaps $P_-(\Delta)\sim |\Delta|^{-\gamma}$ for $\Delta<0$, with $\gamma\approx 0.2$, which again does not vary with $L$. 
    (c-d) Distribution of overlaps and gaps for tanh-networks. The exponents in this case are different: $\theta\approx 0.2$, $\gamma\approx0.16$. \label{fig:overlaps}}
\end{figure*}

We now study the distribution of \emph{gaps} $\Delta<0$ and \emph{overlaps} $\Delta>0$, which play an important role near jamming. Positive $\Delta$'s are associated with unsatisfied patterns --- which increase the loss of the system --- whereas negative $\Delta$'s correspond to satisfied patterns --- which are correctly classified with a margin $\epsilon$ and do not contribute to the loss. The latter offer an important measure not only at the jamming transition, but also in the overparametrized regime, where they signal how much room is left around a minimum of the loss to fit additional patterns. In Fig.~\ref{fig:overlaps} (a,b) we show the two distributions for different depths $L=2,3,5$ (positive $\Delta$'s have been rescaled by $\mathcal{L}^{1/2}$). Remarkably, they behave as power laws for about two decades, $P_+(\Delta/\sqrt{\cal L})\sim (\Delta/\sqrt{\cal L})^\theta$ and $P_-(\Delta) \sim |\Delta|^{-\gamma}$, with novel exponents $\theta\approx0.3$ and $\gamma\approx0.2$ that \mm{appear to} differ from those found for the jamming of particles (which are $\theta\approx0.42311\ldots$ and $\gamma\approx0.41269\ldots$ ). For comparison, in Fig.~\ref{fig:overlaps} (c,d) we show the distribution of the same variables for tanh-networks, that also display power-law behaviors but with different exponents $\theta \approx 0.2$ and $\gamma\approx0.16$.

\begin{figure*}
    \centering
    \setlength{\unitlength}{0.1\textwidth}
    \begin{picture}(10,5)
    \put(0,2.5){\def\svgwidth{0.23\textwidth}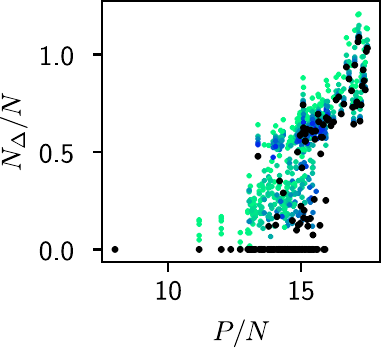}\put(1.3,4.7){(a)}
    \put(2.5,2.55){\def\svgwidth{0.23\textwidth}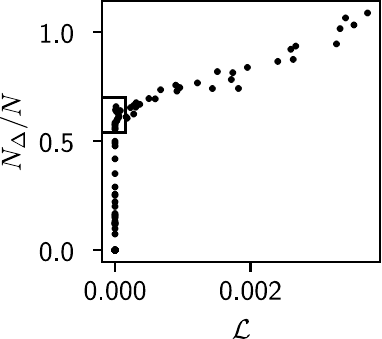}\put(3.8,4.7){(b)}
    \put(5,2.65){\def\svgwidth{0.23\textwidth}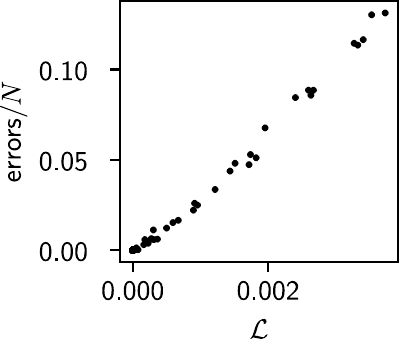}\put(6.35,4.7){(c)}
    \put(7.5,2.65){\def\svgwidth{0.23\textwidth}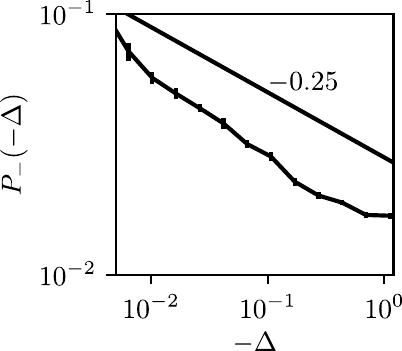}\put(8.8,4.7){(d)}
    \put(0.25,0.05){\def\svgwidth{0.21\textwidth}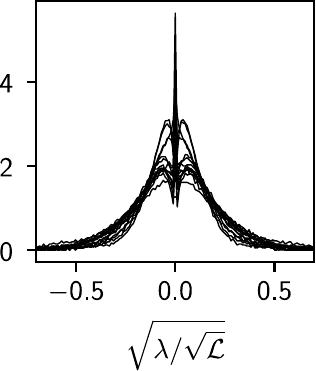}\put(1.3,2.2){(e)}\put(-0.05,1.2){$\mathcal{H}_p$}
    \put(2.75,-0.1){\def\svgwidth{0.19\textwidth}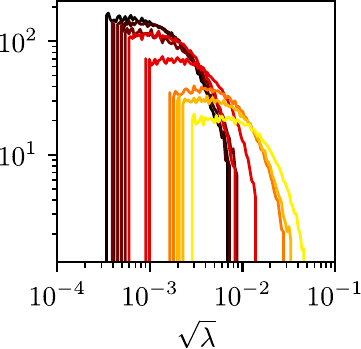}\put(3.8,2.2){(f)}\put(2.45,1.2){$\mathcal{H}_p$}
    \put(5.25,0){\def\svgwidth{0.21\textwidth}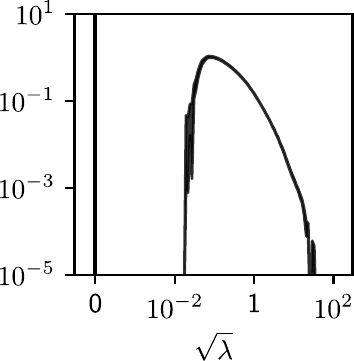}\put(6.3,2.2){(g)}\put(4.95,1.2){$\mathcal{H}_0$}
    \put(7.75,0){\def\svgwidth{0.21\textwidth}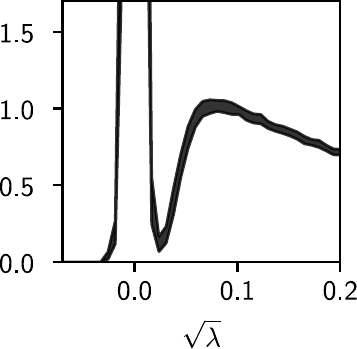}\put(8.8,2.2){(h)}\put(7.45,1.2){$\mathcal{H}$}
    \end{picture}
    \caption{Results with the MNIST dataset, keeping the first 10 PCA components. $d=10$, $h=30$ and $L=5$ ($N= 3900$), varying $P=1,\dots,70k$. (a) The number of unsatisfied patterns $N_\Delta/N$ jumps discontinuously when $r=P/N$ is increased. (b) The same quantity is less noisy when plotted against the loss. (c) The number of misclassified data is a smooth function of the loss. (d) Distribution of the negative gaps ($\Delta<0$), with a tentative exponent $\gamma=0.25$. In the second row (e-h), the Hessian of the runs contained in the rectangle of plot (b) are shown: (e) positive part of the spectrum of $\mathcal{H}_p$, in logarithmic scale; (f) the total spectrum of $\mathcal{H}_p$ appears to be symmetric; (g) the spectrum of $\mathcal{H}_0$ presents a delta function in zero and a gapped continuous spectrum at high frequencies; (h) the spectrum of the total Hessian $\mathcal{H}$ resembles that for random data: the delta function in the spectrum of  $\mathcal{H}_0$ is smeared. 
    }
    \label{fig:mnist}
\end{figure*}

In the case of spheres, the two exponents are related by an inequality that happens to be saturated~\cite{Wyart12,Lerner13a}. The inequality comes from arguments on the stability of jammed packings, and the fact that it is saturated \mm{(which can be proven for certain dynamics \cite{Muller14})} implies that such systems are \emph{marginally stable}: they display an abundance of low-energy excitations and are prone to avalanche dynamics and crackling response when perturbed \cite{Muller14}, a property associated with a hierarchical organization of the loss landscape \cite{Charbonneau14,Franz17b}. The presence of such power laws for deep networks thus suggests they are marginally stable as well, and that the learning dynamics may occur by avalanches where the unsatisfied constraints change by bursts. This will be subject of detailed studies in a future paper.


\section{Image data: MNIST}

We now consider a dataset called MNIST, which consists of a collection of black and white pictures of $28\times28$ pixels depicting handwritten digits from 0 to 9. The labels $y_\mu$ in principle would be the digits themselves ($y_\mu\in\left\{0,\dots,9\right\}$), but to compare more directly with our previous experiments we gathered all the digits into two groups (even and odd numbers) with labels $y_\mu=\pm1$. The architecture of the network is as in the previous sections: the $d$ inputs are fed to a cascade of $L$ fully-connected layers with $h$ neurons, that in the end result in a single scalar output. The loss function used is the hinge loss.

\begin{figure}[ht]
    \centering
    \def\svgwidth{0.7\columnwidth}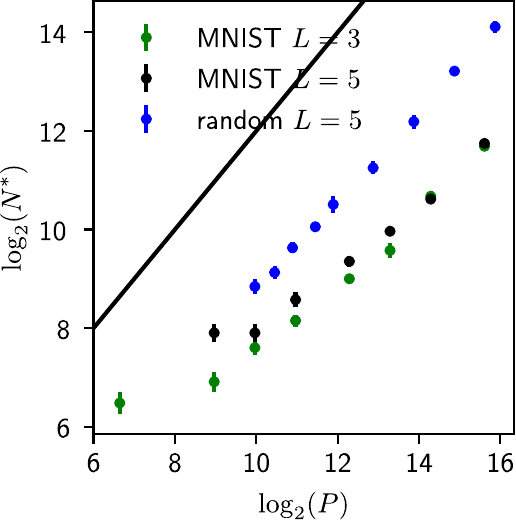
    \caption{Results with the MNIST dataset, keeping the first 10 PCA components (see main text), with $d=10$ and varying $P$ and $h$. The plot shows the number of parameters $N^*$ at the jamming transition. For comparison, we also show the theoretical upper bound (solid curve) and the results found with random data (black points). The maximum number of steps is $2\cdot10^6$.}
    \label{fig:mnist_trans}
\end{figure}

If we kept the original input size of $28\times28=784$ then the majority of the network's weights would be necessarily concentrated in the first layer (the width $h$ cannot be too large in order to be able to compute the Hessian). To avoid this issue, we opted for a reduction of the input size. We performed a principal component analysis (PCA) on the whole dataset and we identified the 10 dimensions that carry the most variance; then we used the components of each image along these directions as a new input of dimension $d=10$. This projection hardly diminishes the performance of the network (we find the generalization accuracy to be larger than $90\%$ at the jamming transition in Fig.~\ref{fig:mnist_trans} for $P\geq 10^4$). 

In Fig.~\ref{fig:mnist} we show that a jamming transition is also found for real data with a discontinuous behavior of $N_\Delta/N$. Fig.~\ref{fig:mnist} (a) shows the number of unsatisfied patterns per parameter $N_\Delta/N$ increasing $P$ at fixed $N$, and in Fig.~\ref{fig:mnist} (b) the same quantity is plotted against the loss. As for random data, the latter is less noisy. In Fig.~\ref{fig:mnist} (c) we show that the number of misclassified data (i.e.\ the number of patterns with $y_\mu f(\mathbf{x}_\mu)<0$) grows smoothly with the loss. These plots depict the same scenario as we found for random data, namely the one presented in Fig.~\ref{fig:errorloss} (a-c), \mm{except for the magnitude of the density of constraints at the transition with } $N_\Delta/N\approx0.5$ rather than $N_\Delta/N\approx0.7$ as observed before. Hence, the number of unsatisfied patterns at the transition is not universal.

Also the spectrum of the Hessian matrix is similar to that of random data. In Fig.~\ref{fig:mnist} (e-h) we show the positive part of the spectrum of $\mathcal{H}_p$, the total spectrum of $\mathcal{H}_p$, the spectrum of $\mathcal{H}_0$ and the spectrum of the total Hessian $\mathcal{H}$, respectively. As with random data: the matrix $\mathcal{H}_p$ has a symmetric spectrum and the matrix $\mathcal{H}_0$ has a finite number of zero modes and a gapped continuous distribution of modes at high energy. The spectrum of the total Hessian is again similar to that of $\mathcal{H}_0$, where the delta function in zero has been smeared.

The distribution of gaps (negative $\Delta$'s) is plotted in Fig.~\ref{fig:mnist} (d), suggesting a power law with an exponent $\gamma=0.25$ that is slightly larger than the value found for random data, $\gamma\approx0.2$. It is unclear whether this difference is significant. We observed that the distribution of overlaps (positive $\Delta$'s) has large sample to sample variations (not shown), and the acquisition of enough statistics to measure it extensively will be done elsewhere.

A key difference between random and structured data however is the location $N^*$ of the transition, shown in Fig.~\ref{fig:mnist_trans} versus the number $P$ of patterns. For a fixed number $P$ of MNIST pictures we ran several simulations with networks of different sizes, and found in this way the lowest value $N^*$ for which all patterns could still be classified correctly. In the figure we present the results for two network architectures of different depths $L=1,3,5$ (the width $h$ was varied in order to control the network size). Key results are  that (i) $N^*$ is essentially independent of depth, especially at larger $P$ and (ii) the minimum number of parameters $N^*$  to fit the data is significantly smaller than for random data, a difference that seems to increase with $P$. The behavior of $N^*$ in the (hypothetical) limit $P\rightarrow \infty$ could be indeed different from the linear scaling of random data: a sub-linear scaling or even a finite asymptotic value are possible alternatives. \mm{More generally, how the data structure affects the location of the transition $N^*(P)$ is an important question for the future.} 




\section{Conclusion}

By slightly changing the loss function --- i.e.\ by considering the hinge loss rather than the commonly used cross entropy, a change that does not degrade performance --- we could recast the problem of minimizing the loss function of deep networks into a constraint satisfaction problem with continuous degrees of freedom. This kind of problem has been abundantly studied in physics, in particular in the context of the jamming of particles, and some theoretical tools developed in that field readily apply to deep networks. In particular from this analogy one predicts a sharp transition as the number of parameters is reduced, separating a region where all constraints can be satisfied (that is, all the data are perfectly fitted) and the loss is zero after learning, and a region where the ratio of the number of unsatisfied constraints to the number of parameters is of order one. This ratio jumps discontinuously at the transition, where it attains a value smaller than one. Near that point, the spectrum of the Hessian is singular, reminiscent of a critical behavior. One key finding is that  deep learning falls into the hypostatic universality class, similar to that of ellispes. We also observe a scaling behavior and new exponents characterizing how well constraints are satisfied or not (through the distributions $P_-(\Delta)$ and $P_+(\Delta)$, respectively). This bears comparison with the known behavior of packings of particles --- where such singularities signal marginality and avalanche-type response --- and of the perceptron (the simplest, shallow, neural network), that lie in the same universality class. Yet there is no theory so far to explain these exponents for deep networks. These results also shed light on \mm{some} aspects of deep learning:

{\it Not getting stuck in poor minima of the loss:} Our analysis supports that in the overparametrized regime, the dynamics does not get stuck in poor minima because the number of constraints to satisfy (data to fit) P is too small to hamper minimization: the system is in an easy satisfiable phase. In particular assuming that a certain operator (namely the matrix $\mathcal{H}_p$) has a fraction of negative eigenvalues (which we could show in the case of the ReLU activation function and random data, and confirm numerically) implies that no poor minima exist if $P/N<P/N^* = \mathcal{O}(1)$. Here $N$ is the number of effective degrees of freedom of the network,
which in all the cases we studied is essentially equal to the number of parameters.
This argument does not rule out the possibility that, with a very poor choice of initial condition,  a poor minimum of the loss can be found. This is the case in particular if the network does not propagate the signal (then $N=1$ in our formalism, independently of the number of parameters). Presumably usual tricks used to train deep networks (batch normalization, residual links, proper weight initialization, ...) ensure that \mm{the sensitivity of the network to its parameters is preserved during training so that} $N$ is indeed similar to the number of parameters, a hypothesis that would be useful to test in a broader setting.

In the under-parametrized phase the network gets stuck at a positive loss, either because the ground state is no longer at zero loss or because the system is trapped in an excited local minimum. The fact that the jamming transition itself depends on the dynamics (as is the case for the jamming of particles) suggests that in the underparametrized case the network is in a local minimum.

{\it Role of depth:} We observed that depth is not helpful to fit random data in fully connected networks: increasing depth and reducing width so that the total number of weights is fixed does not allow to fit the data with less parameters. We have also observed that this finding continues to hold in a realistic case based on MNIST. This may seem to clash with mathematical results, such as \cite{Montufar14,Bianchini14,Raghu16,lee2017ability}, which establish that depth enhances expressivity. However, we tackle the question of expressivity for \emph{realistic} data and learning protocols, which is quite different. Our results, that need confirmation by further studies on a broader range of data, point toward a negative answer for fully connected networks. It may be that the added expressive power of deep networks is only useful for architectures exploiting the symmetry and hierarchy in the data (e.g.\ as in convolutional networks). Alternatively, depth may play a role \mm{in accelerating the learning dynamics}  \cite{shwartz2017opening}.

{\it Reference point for network architectures:} key properties of deep networks, including the learning dynamics and the generalization power, are believed to be affected by the landscape geometry. We have argued that there exists a critical line $N^*(P)$ where the landscape is singular (with both flat and stiff directions), suggesting that it will be a useful reference point to study dynamics and generalization. Concerning the former, our observations suggest that learning near threshold may occur by avalanches, that is, by abrupt changes in the set of data that are correctly classified. In practice, networks are generally trained in the overparametrized regime $N \gg N^*$. It would be interesting to investigate whether the learning dynamics, at intermediate times where many data are not fitted yet, resembles the dynamics near threshold and displays bursts of changes in the constraints. Concerning the latter,  we have studied the effect of jamming on generalization since this article was first written, as appears in \cite{spigler2018jamming,geiger2019scaling}. 

\begin{acknowledgments}
We thank C. Brito, C. Cammarota, T.S. Cohen, S. Franz, Y. LeCun, F. Krzakala, R. Ravasio, P. Urbani and L. Zdeborova for helpful discussions.
This work was partially supported by the grant from the Simons Foundation (\#454935 Giulio Biroli, \#454953 Matthieu Wyart). M.W. thanks the Swiss National Science Foundation for support under Grant No. 200021-165509.\\
\mm{The manuscript ~\cite{franzpre}, which appeared at the same time than ours, shows that the critical properties of the jamming transition} found for
the non-convex perceptron \cite{Franz16} hold more generally in some shallow networks. This universality is an intriguing result. Understanding 
the connection with our findings, which show instead a jamming transition similar to that of ellipses,
is certainly worth future studies.  

\end{acknowledgments}

\bibliography{main.bib}

\appendix

\section{Effective number of degrees of freedom}
\label{app:neff}



Due to several effects discussed in the main text, the function $f(\mathbf{x}; \mathbf{W})$ can effectively depend on less variables that the number of parameters, and thus reduce the dimension of the space spanned by the gradients  $\nabla_\mathbf{W}f(\mathbf{x}; \mathbf{W})$ that enters in the theory. For instance, there could be symmetries that reduce the number of effective degrees of freedom (e.g.\ each ReLU activation function has one of such symmetries, since one can rescale inputs and outputs in such a way that the post-activation is left invariant); another reason could be that a neuron might never activate for all the training data, thus effectively reducing the number of neurons in the network; furthermore, we expect that the network's true dimension would also be reduced if its architecture presents some bottlenecks, is poorly designed or poorly initialized. For example if all biases are too negative on the neurons of one layer in the Relu case, the network does not transmit any signals, leading to $N=1$ and to the possible absence of unstable directions even if the number of parameters is very large.

It is tempting to define the effective dimension by considering the dimension of the space spanned by $\nabla_\mathbf{W}f(\mathbf{x_\mu}; \mathbf{W})$ as $\mu$ varies. This definition is not  practical for small number of samples $P$ however, because this dimension would be bounded by $P$. We can overcome such a problem by considering a neighborhood of each point $\mathbf{x}_\mu$, where the network's function and its gradient can be expanded in the pattern space:
\begin{equation}
    f(\mathbf{x}) \approx f(\mathbf{x}_\mu) + (\mathbf{x} - \mathbf{x}_\mu) \cdot \nabla_{\mathbf{x}} f(\mathbf{x}_\mu),
\end{equation}
\begin{equation}
    \nabla_\mathbf{W} f(\mathbf{x}) \approx \nabla_\mathbf{W} f(\mathbf{x}_\mu) + (\mathbf{x} - \mathbf{x}_\mu) \cdot \nabla_{\mathbf{x}} \nabla_\mathbf{W} f(\mathbf{x}_\mu).
\end{equation}
Varying the pattern $\mu$ and the point $\mathbf{x}$ in the neighborhood of $\mathbf{x}_\mu$, we can build a family $M$ of vectors:
\begin{equation}
    M = \left\{\nabla_\mathbf{W} f(\mathbf{x}_\mu) + (\mathbf{x} - \mathbf{x}_\mu) \cdot \nabla_{\mathbf{x}} \nabla_\mathbf{W} f(\mathbf{x}_\mu)\right\}_{\mu,\mathbf{x}}.
\end{equation}
We then define the effective dimension $N$ as the dimension of  $M$. Because of the linear structure of $M$, it is sufficient to consider, for each $\mu$, only $d+1$ values for $x$, e.g.\ $x-x_\mu=0,\hat{\mathbf{e}}_1,\dots,\hat{\mathbf{e}}_d$, where $\hat{\mathbf{e}}_n$ is the unit vector along the direction $n$. The effective dimension is therefore
\begin{equation}
    N = \mathrm{rk}(G),
\end{equation}
where the elements of the matrix $G$ are defined as
\begin{equation}
    G_{i,\alpha} \equiv \partial_{W_i} f(\mathbf{x}_\mu) + \hat{\mathbf{e}}_n \cdot \nabla_{\hat{\mathbf{e}}_n} \partial_{W_i} f(\mathbf{x}_\mu),
\end{equation}
with $\alpha\equiv(\mu,n)$. The index $n$ ranges from $0$ to $d$, and $\hat{\mathbf{e}}_0 \equiv 0$.

\begin{figure}[htb]
    \centering
    \def\svgwidth{0.7\columnwidth}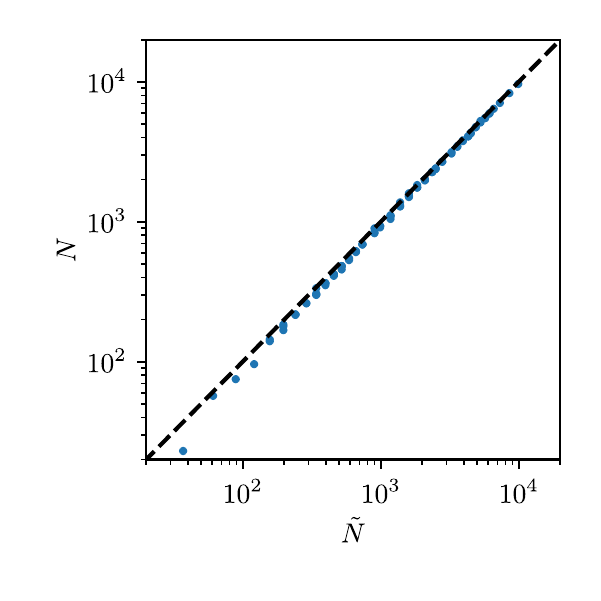
    \caption{Results with the MNIST dataset, keeping the first 10 PCA components. $d=10$ and $L=3$, varying $P$ and $h$. Effective $N$ vs total number of parameters $\tilde{N}$. $N$ is always smaller than $\tilde{N}$ because there is a symmetry per each ReLU-neuron in the network.}
    \label{fig:neff}
\end{figure}

In Fig.~\ref{fig:neff} we show the effective number of parameters $N$ versus the total number of parameters $\tilde{N}$, in the case of a network with $L=3$ layers trained on the first 10 PCA components of the MNIST dataset. There is no noticeable difference between the two quantities: the only reduction is due to the symmetries induced by the ReLU functions (there is one such symmetry per neuron. Indeed the ReLU function $\rho(z) = z \Theta(z)$ satisfies $\Lambda \rho(z/\Lambda) \equiv \rho(z)$.) We observed the same results for random data.

\section{\texorpdfstring{$\mathrm{sp}(H_p)$}{sp(Hp)} is symmetric for ReLu activation function and random data}
\label{app:hpsym}

We consider $\mathcal{H}_p = -\sum_\mu y_\mu \rho\,(\Delta_\mu)\, \hat{\mathcal{H}}_\mu$, where $\hat{\mathcal{H}}_\mu$ is the Hessian of the network function $f(\mathbf{x}_\mu;\mathbf{W})$ and $\rho$ is the Relu function. We want to argue that the spectrum of $\mathcal{H}_p$ is symmetric in the limit of large $N$. 

We do two main hypothesis: First, the trace of any finite power of $\mathcal{H}_p$ is self-averaging (concentrates) with respect to the average over the random data:
$$\frac{1}{N}\mathrm{tr}(\hat{\mathcal{H}_p}^n) =\frac{1}{N}\overline{\mathrm{tr}(\hat{\mathcal{H}_p}^n) }.$$
Second, 
$$
\frac{1}{N}\sum_{\mu_1,\cdots,\mu_n} 
\overline{y_{\mu_1}\rho(\Delta_{\mu_1})\cdots y_{\mu_n}\rho(\Delta_{\mu_n})
\mathrm{tr}(\hat{\mathcal{H}}_{\mu_1}\cdots \hat{\mathcal{H}}_{\mu_n}) }
=$$
$$
\frac{1}{N}\sum_{\mu_1,\cdots,\mu_n} 
\overline{y_{\mu_1}\rho(\Delta_{\mu_1})\cdots y_{\mu_n}\rho(\Delta_{\mu_n})}\overline{
\mathrm{tr}(\hat{\mathcal{H}}_{\mu_1}\cdots \hat{\mathcal{H}}_{\mu_n}) }
$$

The first hypothesis is natural since $\hat{\mathcal{H}_p}$ is a very large random matrix, for which the density of eigenvalues is expected to become a non-fluctuating quantity. The second hypothesis is more tricky: it is natural to assume 
that the trace concentrates, however one also need to 
show that the sub-leading corrections to the self-averaging of the trace 
can be neglected. 

Using these two hypothesis and the result, showed below, 
that 
\begin{equation}\label{eqtra}
    \overline{
\mathrm{tr}(\hat{\mathcal{H}}_{\mu_1}\cdots \hat{\mathcal{H}}_{\mu_n}) }=0
\end{equation}
for all $n$ odds, one can conclude that all odds traces of $\hat{\mathcal{H}_p}$ are zero. This implies that 
the spectrum of $\hat{\mathcal{H}_p}$ is symmetric, more precisely that the fractions of negative and positive eigenvalues are equal.

\begin{figure*}
    \centering
    \def\svgwidth{0.9\textwidth}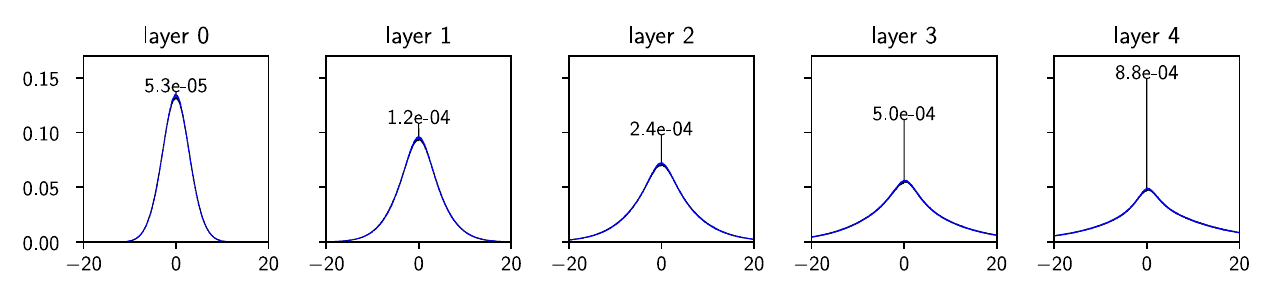
    \caption{\small Density of the pre-activations for each layers with $L=5$ and random data, averaged over all the runs just above the jamming transition with that architecture. Black:  distribution obtained over the training set. Blue: previously unseen random data (the two curves are on top of each other except for the delta in zero). The values indicate the mass of the peak in zero, which is only present when the training set is considered.}
    \label{fig:preactivity}
\end{figure*}

In order to show that the statement (\ref{eqtra}) above holds, let us argue first that $\overline{\mathrm{tr}(\hat{\mathcal{H}}^n_\mu)} = 0$ for any \emph{odd} $n$.
\begin{equation}
    \mathrm{tr}(\hat{\mathcal{H}}^n_\mu) = \sum_{i_1,i_2,\dots,i_n} \hat{\mathcal{H}}^\mu_{i_1,i_2} \hat{\mathcal{H}}^\mu_{i_2,i_3} \cdots \hat{\mathcal{H}}^\mu_{i_n,i_1},
    \label{eq:tracen}
\end{equation}
where the indices $i_1,\dots,i_n$ stand for synapses connecting a pair of neurons (i.e.\ each index is associated with a synaptic weight $W^{(j)}_{\alpha,\beta}$: we are not writing all the explicit indexes for the sake of clarity). The term of the hessian obtained when differentiating with respect to weights $W^{(j)}_{\alpha,\beta}$ and $W^{(k)}_{\gamma,\delta}$ reads
\begin{multline}
    \hat{\mathcal{H}}^{\mu;(jk)}_{\alpha\beta;\gamma\delta} = \sum_{\pi_0,\dots,\pi_{L}} \theta(a^\mu_{L,\pi_{L}})\cdots\theta(a^\mu_{1,\pi_{1}}) x^\mu_{\pi_{0}} \cdot\\\cdot \partial_{W^{(j)}_{\alpha,\beta}} \partial_{W^{(k)}_{\gamma,\delta}} \left[W^{(L+1)}_{\pi_{L}} W^{(L)}_{\pi_{L},\pi_{L-1}} \cdots W^{(1)}_{\pi_{1}\pi_{0}}\right].
    \label{eq:hessianfull}
\end{multline}
where we denoted with $a$ the inputs in the nodes of the network. 
Our argument is based on a symmetry of the problem with random data: changing the sign of the weight of the last layer $W^{(L+1)} \longrightarrow -W^{(L+1)}$ and changing the labels $y_\mu\longrightarrow -y_\mu$ leaves the loss unchanged. 
We will show that this symmetry implies that $\mathrm{tr}(\hat{\mathcal{H}}^n_\mu)$ averaged over the random labels is zero for odd $n$. 

In fact, note that the sum in Eq.\ref{eq:hessianfull} contains a weight per each layer in the network, with the exception of the two layers $j,k$ with respect to which we are deriving. This implies that any element of the hessian matrix where we have not differentiated with respect to the last layer ($j,k < L+1$) is an odd function of the last layer $W^{(L+1)}$, meaning that if $W^{(L+1)} \longrightarrow -W^{(L+1)}$, then the sign of all these Hessian elements is inverted as well.

If in the argument of the sum in Eq.~(\ref{eq:tracen}) there is no index belonging to the last layer, then the whole term changes sign under the transformation $W^{(L+1)} \longrightarrow - W^{(L+1)}$. Suppose now that, on the contrary, there are $m$ terms with one index belonging to the last layer (we need not consider the case of two indices both belonging to the last layer because the corresponding term in the Hessian would be $0$, as one can see in Eq.~(\ref{eq:hessianfull})). For each index equal to $L+1$ (the last layer), there are exactly two terms: $\hat{\mathcal{H}}^\mu_{j,L+1} \hat{\mathcal{H}}^\mu_{L+1,k}$ (for some indexes $j,k$). Since $j,k$ cannot be $L+1$ too, this implies that the number $m$ of terms with an index belonging to the last layer is always even. Consequently, when the sign of $W^{(L+1)}$ is reversed, the argument of the sum in Eq.~(\ref{eq:tracen}) is multiplied by $(-1)^{n-m}$ (once for each term \emph{without} an index belonging to the last layer), which is equal to $-1$ if $n$ is odd.
The same symmetry can be used to show that a matrix made of an odd product of matrices $\hat{\mathcal{H}}_\mu$, such as $\hat{\mathcal{H}}_\mu \hat{\mathcal{H}}_{\mu'}\hat{\mathcal{H}}_{\mu''}$, must also have a symmetric spectrum,
concluding our argument. 

\section{Density of pre-activations for ReLU activation functions}
\label{app:zeros}

The densities of pre-activation (i.e. the value of the neurons before applying the activation function) is shown in Fig.~\ref{fig:preactivity} for random data. It contains a delta distribution in zero. The number $N_c$ of pre-activations equal to zero when feeding a network $L=5$ all its random dataset is $N_c\approx 0.21 N$, corresponding to the number of directions in phase space where cusps are present in the loss function. For MNIST data we find $N_c\approx 0.19N$. By taking $L=2$ and random data we find $N_c\approx 0.25N$. In these directions, stability can be achieved even if the hessian would indicate an instability. For this reason, instead of $N_-$ in Equation~\ref{4bis} one should use $N/2-N_c\approx0.25 N$.

\end{document}